# Experimental heating of complex organic matter at Titan's interior conditions supports contributions to atmospheric $N_2$ and $CH_4$

**K. E. Miller[1*], D. I. Foustoukos[2], G. Cody[2], and C. M. O'D. Alexander[2]**

[1]Southwest Research Institute, San Antonio, TX. [2]Earth and Planets Laboratory, Carnegie Institution for Science, 5241 Broad Branch Road, Washington, D.C. 20015

[*]Corresponding author: Kelly Miller (kelly.miller@swri.org)

**Abstract**

Titan's abundant atmospheric $N_2$ and $CH_4$ gases are notable characteristics of the moon that may help constrain its origins and evolution. Previous work suggests that atmospheric $CH_4$ is lost on geologically short timescales and may be replenished from an interior source. Isotopic and noble gas constraints indicate that $N_2$ may derive from a mixture of $NH_3$ ice and heating of organic matter. Here, we report experimental results from hydrothermal alteration of insoluble organic matter from the Murchison meteorite and analog insoluble organic matter at temperatures and pressures that are relevant to Titan's interior. Our results indicate both $CH_4$ and $CO_2$ are formed, with the ratio between the two depending on a multitude of factors, particularly temperature and, to a lesser degree, the dielectric constant of water and carbonyl abundance in the starting material. Sufficient $CH_4$ is produced to source Titan's atmospheric reservoir if temperatures are greater than 250 °C. Nitrogen is volatilized, primarily in the form of $NH_3$, in sufficient abundances to source at least 50 % of Titan's atmospheric $N_2$. The isotopic characteristics of volatilized material relative to the starting organics are consistent with current constraints for the nature of the accreted complex organics and Titan's evolved atmosphere.

# 1 Introduction

The origin of Titan's thick atmosphere may provide key clues not only to the formation and evolution of Titan but also to the formation conditions of the Saturnian system. The present-day atmosphere of Titan has a base pressure of approximately 1.5 bar and is dominated by $N_2$ gas. The presence of between 1.5 mole% and 5.7 mole% $CH_4$ is, however, critical to the maintenance of the relatively warm temperatures required for $N_2$ to remain in the gas phase (Charnay et al. 2014; Lorenz et al., 1997). Measurements by the Cassini Huygens probe indicate that there are trace quantities of $^{36}$Ar and $^{40}$Ar in Titan's atmosphere, while Xe and Kr were below detection with upper limit mole fractions of $1\times10^{-8}$ (Niemann et al., 2010). Measurement of the $^{14}$N/$^{15}$N ratio in $N_2$ (~168; Niemann et al., 2010), and the $^{12}$C/$^{13}$C (91.4±1.5; Niemann et al., 2005, 2010) and D/H ([1.35±0.3]$\times10^{-4}$; Niemann et al., 2010) ratios in $CH_4$ are also recognized as key constraints in the current understanding of the formation of Titan's atmosphere (Mandt et al., 2009, 2012; Nixon et al., 2012).

The primary accretion of $N_2$ at Titan as a volatile ice was initially ruled out by the limit set on the abundance of Ne relative to $N_2$ from the Voyager UV, IRIS, and radio experiments (Strobel and Shemansky, 1982). These measurements indicated strongly subsolar Ne relative to N. This could be explained by accretion of $N_2$ as a clathrate, to the exclusion of Ne (Owen, 1982); but if true, then a solar Ar/$N_2$ ratio would be expected. Measurements by the Huygens probe GC-MS experiment, however, showed that $^{36}$Ar/$N_2$ is subsolar by a factor of approximately $10^5$ (Niemann et al., 2010), ruling out accretion of $N_2$ ices as the primary source for Titan's atmospheric $N_2$. Secondary production of $N_2$ from N-bearing precursors must therefore be a significant process at Titan.

One possible precursor form is gaseous $NH_3$, which is the thermodynamically favored volatile form of N in disk environments. Kinetic inhibition of the formation of $NH_3$ from $N_2$ (Lewis and Prinn, 1980) may be overcome in a warm, well-mixed Saturnian subnebula (Prinn and Fegley, 1981). However, orbital evolution of giant planet-satellite systems may instead point towards formation in a gas-starved disk (e.g. Canup and Ward, 2006). Under such conditions, conversion of gaseous $N_2$ to $NH_3$ is kinetically inhibited at Titan's distance from Saturn (Alibert and Mousis, 2007), favoring $NH_3$ ice as the N source.

High $^{15}N/^{14}N$ ratios in cometary $NH_3$ ice were predicted on the basis of Titan's atmospheric composition (Sekine et al., 2011). Later measurements of $^{15}NH_2/^{14}NH_2$ in comets (Rousselot et al., 2013; Shinnaka et al., 2014, 2016) confirmed this prediction. Modeling of atmospheric loss processes at Titan indicate that the evolution of the $^{15}N/^{14}N$ ratio of the atmosphere over the lifetime of the solar system points towards cometary $NH_3$ ice as a significant source of N for Titan's atmosphere (Erkaev et al., 2021; Mandt et al., 2009, 2014). Hydrodynamic escape and sputtering were found to be inefficient at N isotopic fractionation, while Jeans escape would act too slowly to isotopically fractionate N within the lifetime of the solar system (Mandt et al., 2014). When the effects of the evolution of the extreme ultraviolet (EUV) flux from the young sun on photochemical loss processes are taken into account, Titan's initial $^{15}N/^{14}N$ ratio is estimated to have been $(5.8-6.0)\times10^{-3}$ (Erkaev et al., 2021), where the cometary value for $NH_2$ is $^{15}N/^{14}N = (7.1-7.7)\times10^{-3}$ independent of cometary dynamic family (Shinnaka et al., 2016).

Accreted $NH_3$ ices may be converted to $N_2$ via photolytic processes and may yield up to 10 bars of $N_2$ in 30 Myr in an $NH_3$ - $CH_4$ primary atmosphere (Atreya et al., 1978, 2010). However, such processes are only effective between 150 K and 250 K, which would require greenhouse warming from a 0.45 bar $CH_4$ - $H_2$ atmosphere prior to the transformation of $NH_3$ to $N_2$ (Atreya et al., 1978). For comparison, the present-day partial pressure of $CH_4$ in Titan's bulk atmosphere is less than 0.1 bar (Niemann et al., 2010), but significant $CH_4$ loss over the lifetime of the atmosphere is expected (Yung et al., 1984). As an alternative to photolytic conversion, $NH_3$ may be oxidized to $N_2$ via impact processes. Experimental work has demonstrated the production of $N_2$ under relevant impact conditions (McKay et al., 1988; Sekine et al., 2011).

However, the impact erosion of Titan's atmosphere is highly effective (Marounina et al., 2015) and may limit the formation of an $N_2$ atmosphere by this process.

Interior processing provides a third possible mechanism for the formation of secondary $N_2$ from $NH_3$ or other accreted forms. Episodic outgassing from Titan's interior has been posited as a mechanism for replenishing $CH_4$ in Titan's atmosphere (Tobie et al., 2006), which is consumed by photochemical processes on timescales of 10 Myr at present rates (Wilson and Atreya, 2004; Yung et al., 1984). Geothermal modeling of Titan's interior suggests that temperatures exceed 400 K within 100 km of the water-rock interface (Castillo-Rogez and Lunine, 2010; Reynard and Sotin, 2023). Endogenic formation of $CH_4$ from C grains or $CO_2$ (Atreya et al., 2006) and of $N_2$ from $NH_3$ (Glein et al., 2009) may occur in a warm interior. Outgassing may also contribute noble gases to Titan's atmosphere. Assumptions about the outgassing efficiency of noble gases, $N_2$, and $CH_4$, based on geochemical behavior including their saturation vapor pressures, suggest that the bulk of Titan's atmosphere may form via outgassing (Glein, 2015).

Assuming chondritic abundances of insoluble organic matter (IOM; up to 4 wt.% in chondrites, e.g. Alexander et al. 2007) in Titan's building blocks, organic contributions to Titan's volatile reservoir from heating in the interior are probable, but likely insufficient to serve as the dominant source (Tobie et al., 2012). Measurements from comet Halley and comet 67P/Churyumov-Gerasimenko suggest that IOM-like material may be a dominant component of outer solar system bodies, though, with abundances on the order of 45 wt.% of the refractory material (Bardyn et al., 2017; Greenberg, 1986; Kissel et al., 1986; Kissel and Krueger, 1987; McDonnell et al., 1986). In this case, the potential for production of atmospheric $N_2$ from IOM-like source material increases significantly, such that production of 50 % of Titan's N atmosphere from hydrothermally degraded IOM is theoretically consistent with both the abundance of N available in such a source and the likely isotopic and noble gas composition of such a source (Miller et al., 2019). The model of Miller et al. (2019) draws on previously published anhydrous pyrolysis experiments on isolated IOM at ambient pressures to estimate outgassing abundances. Additional experimental work on whole rock samples confirms the production of $CH_4$, CO, and $CO_2$ from carbonaceous chondrites at high temperatures and ambient pressures (Thompson et al., 2022). Titan's present-day water-rock interface is at a pressure of approximately 8 kbar (Sotin et al., 2021), and pressure and temperature may both influence volatilization of complex organic matter (Mi et al. 2014). Here, we examine the role of temperature and pressure on the composition and abundance of volatile material derived from complex organic matter, including C-bearing gases as well as production of $NH_3$, and consider the implications for the origins of Titan's volatiles.

## 2 Materials and Methods

### 2.1 Synthesis of syn-IOM

Analog IOM ("syn-IOM") was synthesized in batches following established protocols (Foustoukos et al., 2021). These organic residues have been shown to exhibit similar chemical and molecular compositions and structures to chondritic IOM (Foustoukos et al., 2021). Briefly, approximately 170 mg of dextrose ($\delta^{13}C$ = -9.6 ‰) and 1 mL of ~ 1 M $NH_4Cl_{(aq)}$ ($\delta^{15}N$ = 0.33 ‰, $\delta D$ = -77 ‰) in aqueous solution ($\delta D$ = -250 ‰ or -100 ‰) were added to a borosilicate glass tube. The tube was flame sealed under atmosphere, and then heated to a temperature of 150 °C to 250 °C, where it was held for over 100 hours. The dextrose polymerizes, and the C/N ratio of the final product is controlled by the concentration of $NH_4^+$ in solution. Upon quenching, the vials are transferred to a freezer overnight. The vials are broken open while frozen, and the residue is retrieved and dried under $N_2$. The syn-IOM is rinsed multiple times with deionized water, followed by a final ethanol rinse. The initial composition of syn-IOM and IOM materials used is summarized in Table 1.

### 2.2 Extracted Murchison IOM

Murchison IOM used was a combination of material extracted from the bulk aliquots (A&B) of Murchison using the CsF-HF protocol as described in Cody et al. (2024). Briefly, powdered meteorite samples were first digested with 2 N HCl, before rinsing with Milli-Q water and dioxane. The resulting residue was treated with a combination of aqueous CsF-HF solution and dioxane, which leads to two immiscible layers. Iterative agitation and centrifuging of the sample leads to accumulation of the separated IOM at the boundary between the two layers. The IOM is removed by pipette, rinsed successively with 2N HCl, Milli-Q water, and dioxane, and then dried at 30-50 °C.

Table 1. Starting Composition of Murchison IOM and syn-IOM. Sample names are related to the H/C ratio; note that N/C and O/C also vary.

| Sample | Type | H/C (at.) | N/C (at.) | O/C (at.) | C, wt.% | δ13C, ‰ | H, wt.% | δD, ‰ | N, wt.% | δ15N, ‰ | O, wt.% | δ¹⁸O |
|---|---|---|---|---|---|---|---|---|---|---|---|---|
| HC083 | Syn-IOM | 0.83±0.04 | 0.073±0.002 | 0.38±0.01 | 54.5±0.2 | -9.7±0.0 | 3.79±0.18 | 32±10 | 4.66±0.13 | 0.55±0.38 | 27.3±0.5 | 24.9±0.3 |
| HC095 | Syn-IOM | 0.95±0.03 | 0.057±0.001 | | 58.1±1.0 | -10.2±0.2 | 4.59±0.10 | 55±15 | 3.88±0.03 | 0.93±0.17 | | |
| HC091 | Syn-IOM | 0.91±0.03 | 0.047±0.001 | | 59.1±0.7 | -10.1±0.0 | 4.47±0.14 | 36±10 | 3.24±0.03 | 0.60±0.08 | | |
| HC096 | Syn-IOM | 0.96±0.05 | 0.057±0.000 | | 67.3±0.2 | -10.4 | 5.41±0.26 | -87 | 4.44±0.02 | 1.35 | | |
| HC113 | Syn-IOM | 1.13±0.03 | 0.100±0.006 | 0.27±0.01 | 55.1±1.2 | -10.4±0.5 | 5.19±0.03 | -37±4 | 6.4±0.4 | 2.45±0.33 | 19.5±0.3 | 39.2±0.5 |
| Murchison[*] | IOM | 0.70±0.03 | 0.036±0.001 | 0.18±0.00[**] | 59.9±1.1 | -18.4±0.1 | 3.49±0.12 | 795±6 | 2.48±0.04 | 4.35±0.14 | 16.2±0.3[**] | 13.2±0.6[**] |

[*]Data give the average for IOM from Murchison A and Murchison B from Cody et al. (2024)
[**]Value taken from Alexander et al. (2007)

### 2.3   Heating experiments

Experiments at 1 kbar and 3 kbar were conducted using cold seal pressure vessels at the Carnegie Institution for Science's Earth and Planets Laboratory, while experiments at 10 kbar were performed using a piston cylinder apparatus. The adopted protocols followed those described in Foustoukos (2012) and Foustoukos and Mysen (2015). Briefly, the experiments were performed in Au capsules that were pre-treated with an acid wash to remove contaminants. For each syn-IOM experiment, approximately 20 mg of syn-IOM was loaded into a ~10 mm Au tube that had been sealed at one end. For the Murchison IOM experiments, approximately 10 mg of IOM were used for each aliquot. A mass of water that was approximately equal in mass to the IOM/syn-IOM aliquot was added to the Au tube, which was then welded closed using a PUK 3s professional plus arc welder. The tubes were weighed before and after sealing to ensure that the welding did not cause loss of water. Water was not specifically degassed to remove $O_2$ prior to sealing, and we estimate based on atmospheric composition that at most ~$4\times10^{-7}$ moles of $O_2$ may have been included in the experimental charges, equivalent to ~$2\times10^{-5}$ mmole $O_2$/mg organics. The initial presence of atmospheric $O_2$ in the experimental charges may have resulted in consumption of product $H_2$ to form water, or in oxidation of C to form $CO_2$. However, the abundance of atmospheric $O_2$ initially present in the experimental charges is negligible compared to the abundance of $CO_2$ produced.

For all experiments, the reaction times at the targeted temperature and pressure conditions (Table 2) were 48 hours. After quenching, the capsules were recovered and placed in glass vials with a blue butyl rubber stopper and sealed with an aluminum crimped cap. The vials were subsequently evacuated and backfilled to atmospheric pressure with Ar. To each vial was added 1 mL of 0.1 N HCl to maintain acidic conditions with the aim of converting all dissolved

C and N to $CO_{2(g)}$ and $NH_4^+$, respectively. The Au capsules were then punctured with a steel needle to release the contents.

### 2.4   Gas Analysis

Volatile samples from the headspace were analyzed by a Shimadzu GC-8A gas chromatograph (GC) equipped with a thermal conductivity detector (TCD) and a Carboxen-1010 Plot/Silica Gel column. Calibration was performed via replicate injections of a mixture of 1% $CO_2$, CO, $H_2$, $O_2$, and $CH_4$ with $N_2$ for the balance (Matheson Gas, Item No. GMT10403TC, Lot No. 109-56-13398). The detection limit for these volatiles is 1–5 µmol/kg (Foustoukos et al., 2011). For each measurement, 1 mL of gas at atmospheric pressure was injected into the chromatograph. The measurements were made at a constant temperature of 120 °C, with a TCD current of 60 mA and Ar as the carrier gas.

To verify that headspace gas was representative and dissolution of gas in the sample vial fluid was not significant, we estimated the total pressure in the vial with the product gas (up to 100 bar) and examined the solubility of $CO_2$ (Diamond and Akinfiev, 2003) and $CH_4$ (Duan and Mao, 2006) at 25°C up to that pressure. We find that dissolved $CO_2$ may add at most 9% to the reported abundances, and dissolved $CH_4$ may add at most 0.6%. While product gases dissolved in the fluid are not directly accounted for, their amount is negligible.

### 2.5   Bulk Analyses of Solids

The elemental and isotopic analyses of the organic residues were made with (i) a Thermo Scientific Delta V[plus] mass spectrometer connected to a Carlo Erba (NA 2500) elemental analyzer (CE/EA) via a Conflo III interface for C and N analyses, (ii) a Thermo Finnigan Delta

Plus XL mass spectrometer connected to a Thermo Finnigan Thermal Conversion elemental analyzer (TC/EA) via a Conflo III interface for H analyses, and an iii) Thermo Scientific Delta Q mass spectrometer connected to an EA - Isolink via a Conflo IV interface for H and O analyses. The Conflo III/IV interfaces facilitate the introduction of the $H_2$, $N_2$, $CO_2$, and CO reference gases. In the Delta Plus XL, a dual inlet system facilitates the introduction of the $H_2$ reference gas. An $H_3^+$ correction was determined and applied to the H measurements (Sessions, 2001). We used in-house standards to normalize and correct the data at regular intervals to monitor the accuracy and precision of the measured isotopic ratios and elemental compositions throughout the runs. These in-house standards, which included both gases and solid materials, have been calibrated against international and other certified standards (Standard Mean Ocean Water or SMOW, National Bureau of Standards-22, IAEA-601, Pee Dee Belemnite, and air) from the Isoanalytical Laboratory, the U.S. Geological Survey, the National Bureau of Standards, and the Oztech Trading Company. The reported uncertainties for the elemental and isotopic analyses correspond to the highest 1σ deviations attained from either the replicate analyses of distinct subsamples or the internal standards.

The experimental solid products were washed several times with milliQ water and ethanol, before being dried down at < 50 °C under vacuum. The solid residues were weighed into foil capsules (Ag for H and O, Sn for C and N). Before H analysis, the solid samples were transferred to a zero-blank autosampler and flushed with dry He for at least one hour. Typical sample sizes were 0.3 mg. For the H analyses, replicate samples were analyzed sequentially to check for both sample heterogeneity and small memory effects that occur. Blanks were run before and after samples, again to reduce any memory effects. For example, the H abundances of the duplicate samples generally differed by ≤ 5 % of their absolute values, and the δD values differed by a median of 4 ‰ (the δ notation stands for the deviation of a sample ratio from a standard ratio in parts per thousand, δ (‰) = $(R_{smp}/R_{std} -1)$ x 1000, and in this case R = D/H). The isotopic ratios are reported in ‰ relative to VSMOW (O, H), Pee Dee Belemnite (C) and standard air (N). There is no memory effect for C, N and O analyses.

## 2.6 $NH_4^+$ Analysis

Ammonium concentrations were determined with the indophenol blue method via spectrophotometry (Tecan Inifinite M200 spectrophotometer) at 570 nm (Pérez-Rodríguez et al., 2017). The reagent kit utilized was from Abcam (#102509). Spectrophotometric measurements were performed as duplicates for each sample and reported as average values. The estimated uncertainty (2σ) for the $NH_4^+$ analysis is less than 2%.

To extract $NH_4^+$ from the sample solutions (3 mL), $NH_4^+$ was converted to $NH_{3(g)}$ using a pH-buffered $Na_2B_4O_7$-NaOH solution (1 mL, pH = 12.7), and then diffused into diffusion packets (e.g. Holmes et al., 1998; Pérez-Rodríguez et al., 2017; Sigman et al., 1997). The diffusion process occurred in pre-evacuated vials to enhance the purity and partial pressure of $NH_{3(g)}$ in the headspace. Diffusion packets were made with 1 cm diameter glass fiber disks grade GF/D filters, pre-acidified with 50 μL of 2 M sulfuric acid (4 N) and sandwiched between 2.5-cm diameter, 10-μm pore size Teflon membranes. Samples were incubated for 7 days at 60 °C. After removing the diffusion packets, the GF/D disks were freeze-dried. Every set of analyses was cross-referenced to standard $NH_4^+$ solutions ($δ^{15}N$ = 17 ‰) with concentrations ranging from 1.9 mM to 0.019 mM. The minimum $NH_4^+$ concentration for $δ^{15}N$ analysis is 0.019 mM.

## 3 Results

### 3.1 Headspace gas

The results from GC analyses of the headspace volatiles are summarized in Table 2. The primary constituents identified via GC analysis included $CH_4$, $CO_2$, $H_2$, and CO (Figure 1), as well as small abundances with an elution time consistent with $N_2$ or air. Overall, the abundances are consistently reproduced across different replicates and syn-IOM starting samples, suggesting that the observed results and trends may be broadly applicable for complex organic matter. We are unable to exclude the possibility that contaminant air was introduced at trace quantities during sample preparation or subsequent work-up, and therefore cannot conclude from these experiments that what is likely $N_2$ was produced via hydrous pyrolysis of the samples. The levels of $N_2$ determined by GC analysis were significantly smaller (~ 10% by volume) than those expected under 1 atmosphere partial pressure, suggesting that atmospheric contamination was minor.

To assess whether experimental charges reached equilibrium, we compared the logarithms of the reaction quotient $Q$ and the equilibrium constant $K$ for the reaction

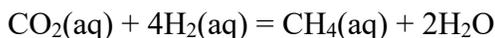

$$CO_2(aq) + 4H_2(aq) = CH_4(aq) + 2H_2O$$

The activities of $CH_4$, $CO_2$, and $H_2$ are taken from the experimental results to calculate $\log(Q)$. We used the CHNOSZ package (Dick, 2019) with the Deep Earth Water model (Sverjensky et al., 2014) implemented for water to calculate $\log(K)$ at our experimental conditions (250 °C, 350 °C, 500 °C; 1 kbar, 3 kbar, 10 kbar). With the possible exception of experiment 350-1-1, our results (Table 2) suggest that the experiments did not achieve equilibrium.

Table 2. Gases measured after heating of experimental charges. Number of detections is given in parentheses after the mean and standard deviation. The log(Q) and log(K) values for reduction of $CO_2$ to $CH_4$ are also given as described in the text.

| Experiment | T, °C | P, kbar | Sample | $H_2$, mmoles/mg sample | CO, mmoles/mg sample | $CH_4$, mmoles/mg sample | $CO_2$, mmoles/mg sample | Air, mmoles/mg sample | log(Q) | Log(K) |
|---|---|---|---|---|---|---|---|---|---|---|
| 250#3 | 250 | 1 | HC095 | $(4.59\pm4.08)\times10^{-7}$ (3) | n.d. | n.d. | $(4.57\pm1.75)\times10^{-5}$ (3) | $(5.84\pm0.79)\times10^{-5}$ (4) | | 14.6 |
| 250-1-1-Mar2023 | 250 | 1 | HC113 | $(1.95\pm0.63)\times10^{-7}$ (4) | $(1.48\pm2.95)\times10^{-8}$ (4) | $(4.04\pm0.31)\times10^{-6}$ (4) | $(1.09\pm0.08)\times10^{-3}$ (4) | $(5.06\pm1.32)\times10^{-5}$ (4) | 12.9 | 14.6 |
| 250-1-2-Mar2023 | 250 | 1 | HC113 | $(9.56\pm0.55)\times10^{-7}$ (3) | n.d. | $(5.27\pm0.40)\times10^{-6}$ (3) | $(1.19\pm0.06)\times10^{-3}$ (3) | $(4.46\pm1.10)\times10^{-5}$ (3) | 10.3 | 14.6 |
| 250#2 | 250 | 3 | HC096 | $(1.27\pm0.31)\times10^{-6}$ (3) | $3.49\times10^{-7}$ (1) | $(8.83\pm1.00)\times10^{-7}$ (3) | $(1.85\pm0.13)\times10^{-4}$ (3) | $(1.10\pm0.43)\times10^{-5}$ (2) | 8.7 | 15.9 |
| 250-3-1-Mar2023 | 250 | 3 | HC113 | $(8.74\pm0.48)\times10^{-7}$ (3) | n.d. | $(3.47\pm0.10)\times10^{-6}$ (3) | $(1.11\pm0.04)\times10^{-3}$ (3) | $(9.54\pm0.88)\times10^{-5}$ (3) | 10.7 | 15.9 |
| 250-3-2-Mar2023 | 250 | 3 | HC113 | $(1.73\pm0.13)\times10^{-6}$ (3) | n.d. | $(4.76\pm0.19)\times10^{-6}$ (3) | $(1.24\pm0.04)\times10^{-3}$ (3) | $(5.26\pm0.70)\times10^{-5}$ (3) | 8.9 | 15.9 |
| 350_1_1 | 350 | 1 | HC083 | $(4.24\pm0.53)\times10^{-6}$ (2) | n.d. | $(1.77\pm0.04)\times10^{-4}$ (2) | $(4.01\pm0.12)\times10^{-3}$ (2) | $(4.11\pm0.78)\times10^{-5}$ (2) | 8.8 | 9.2 |
| 350_1_2 | 350 | 1 | HC096 | $(1.49\pm0.04)\times10^{-6}$ (2) | n.d. | $(1.04\pm0.07)\times10^{-4}$ (2) | $(2.32\pm0.11)\times10^{-3}$ (2) | $(6.84\pm1.44)\times10^{-5}$ (2) | 10.8 | 9.2 |
| 350_3_1 | 350 | 3 | HC083 | $(4.31\pm0.27)\times10^{-6}$ (2) | n.d. | $(1.46\pm0.10)\times10^{-4}$ (2) | $(3.95\pm0.19)\times10^{-3}$ (2) | $(7.58\pm1.41)\times10^{-5}$ (2) | 8.9 | 10.3 |
| 350_3_2 | 350 | 3 | HC096 | $(1.20\pm0.0392)\times10^{-5}$ (2) | n.d. | $(6.90\pm0.15)\times10^{-5}$ (2) | $(2.17\pm0.03)\times10^{-3}$ (2) | $(4.37\pm0.61)\times10^{-5}$ (2) | 6.7 | 10.3 |
| 350-3-M | 350 | 3 | Murchison | $(6.00\pm0.09)\times10^{-6}$ (3) | n.d. | $(1.84\pm0.12)\times10^{-4}$ (3) | $(3.04\pm0.19)\times10^{-3}$ (3) | $(9.38\pm1.56)\times10^{-5}$ (3) | 8.3 | 10.3 |
| 350_10_1 | 350 | 10 | HC083 | $(6.59\pm0.46)\times10^{-6}$ (2) | n.d. | $(1.03\pm0.11)\times10^{-4}$ (2) | $(4.36\pm0.16)\times10^{-3}$ (2) | $(8.97\pm1.08)\times10^{-5}$ (2) | 8.2 | 13.7 |
| 350-10-Mar2023 | 350 | 10 | HC113 | $(1.12\pm0.11)\times10^{-5}$ (3) | n.d. | $(7.26\pm0.63)\times10^{-4}$ (3) | $(2.35\pm0.19)\times10^{-3}$ (3) | $(1.00\pm0.174)\times10^{-4}$ (3) | 8.7 | 13.7 |
| 350-10-M | 350 | 10 | Murchison | $(2.58\pm0.20)\times10^{-5}$ (3) | n.d. | $(2.83\pm0.20)\times10^{-4}$ (3) | $(2.83\pm0.16)\times10^{-3}$ (3) | $(6.71\pm2.01)\times10^{-5}$ (3) | 5.5 | 13.7 |
| 500#1 | 500 | 1 | HC096 | $(3.99\pm0.31)\times10^{-5}$ (4) | $(1.71\pm2.41)\times10^{-7}$ (2) | $(2.64\pm0.17)\times10^{-3}$ (4) | $(1.11\pm0.19)\times10^{-3}$ (4) | $(1.66\pm0.33)\times10^{-5}$ (2) | 5.4 | 2.3 |
| 500#3 | 500 | 1 | HC091 | $(7.55\pm4.32)\times10^{-6}$ (3) | n.d. | $(5.15\pm4.92)\times10^{-4}$ (3) | $(7.24\pm6.64)\times10^{-4}$ (3) | $(5.72\pm0.69)\times10^{-5}$ (2) | 8.7 | 2.3 |

| | | | | | | | | | |
|---|---|---|---|---|---|---|---|---|---|
| 500#2 | 500 | 3 | HC096 | $(6.78\pm0.49)\times10^{-6}$ (3) | $1.13\times10^{-6}$ (1) | $(2.60\pm0.06)\times10^{-3}$ (3) | $(9.49\pm0.87)\times10^{-4}$ (3) | $(3.87\pm0.26)\times10^{-5}$ (2) | 8.5 | 3.9 |
| 500-3-1-Mar2023 | 500 | 3 | HC113 | $(7.84\pm0.25)\times10^{-6}$ (3) | n.d. | $(2.59\pm0.06)\times10^{-3}$ (3) | $(2.22\pm0.04)\times10^{-3}$ (3) | $(4.23\pm0.38)\times10^{-5}$ (3) | 8.9 | 3.9 |
| 500-3-2-Mar2023 | 500 | 3 | HC113 | $(1.32\pm0.08)\times10^{-5}$ (3) | n.d. | $(3.25\pm0.15)\times10^{-3}$ (3) | $(2.81\pm0.11)\times10^{-3}$ (3) | $(2.83\pm1.08)\times10^{-5}$ (3) | 8.4 | 3.9 |
| 500-3-M | 500 | 3 | Murchison | $(3.00\pm0.09)\times10^{-5}$ (3) | n.d. | $(1.57\pm0.08)\times10^{-3}$ (3) | $(3.43\pm0.14)\times10^{-3}$ (3) | $(1.01\pm0.17)\times10^{-4}$ (3) | 6.5 | 3.9 |
| 500#4 | 500 | 10 | HC096 | $(2.44\pm0.26)\times10^{-5}$ (3) | $4.78\times10^{-7}$ (1) | $(3.18\pm0.15)\times10^{-3}$ (3) | $(1.01\pm0.08)\times10^{-3}$ (3) | $(3.11\pm0.35)\times10^{-5}$ (2) | 7.6 | 6.6 |
| 500-10 | 500 | 10 | HC113 | $(3.04\pm0.14)\times10^{-5}$ (4) | n.d. | $(3.40\pm0.15)\times10^{-3}$ (4) | $(2.55\pm0.09)\times10^{-3}$ (4) | $(3.30\pm0.71)\times10^{-5}$ (4) | 7.5 | 6.6 |

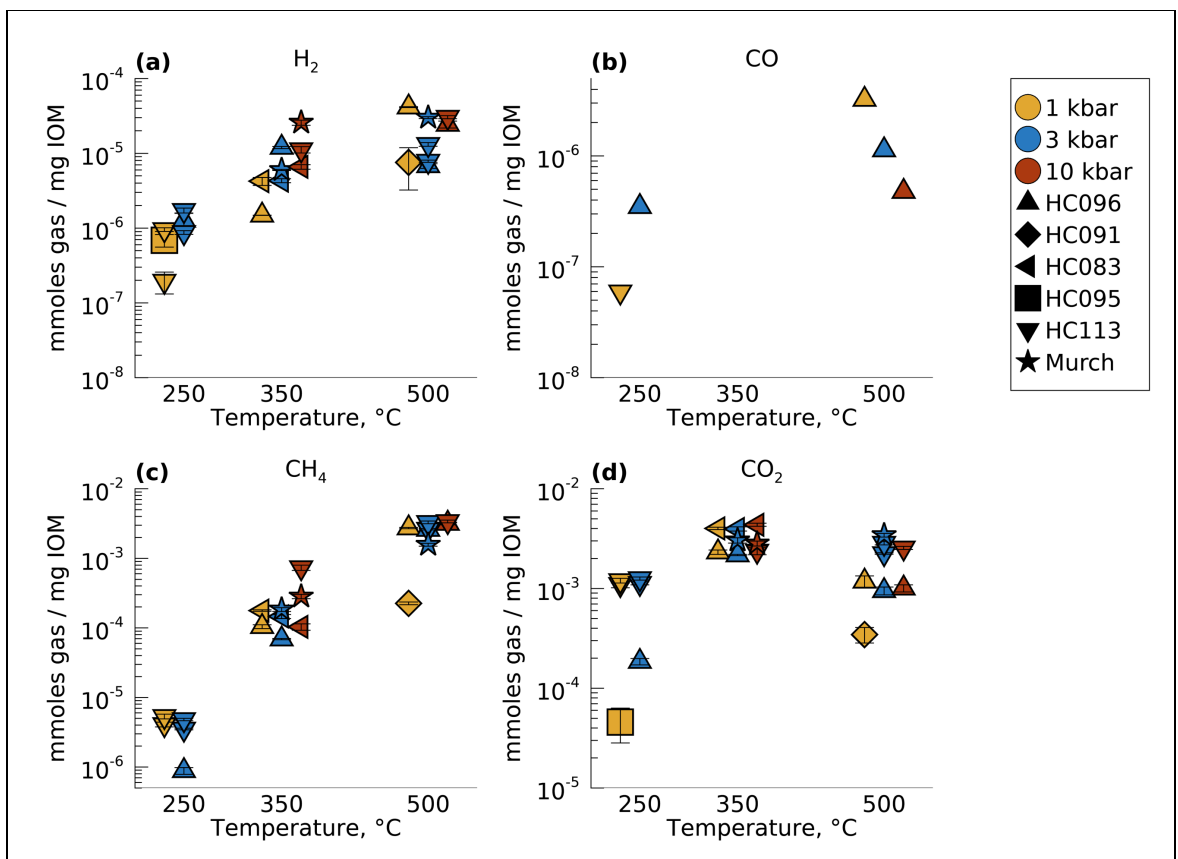

Figure 1. Production of COH volatiles for all experimental charges: (a) $H_2$, (b) CO, (c) $CH_4$, and (d) $CO_2$. The temperature at which experiments were conducted is given on the x-axis, and the color of the symbol gives the pressure. The shape of the symbol corresponds to the syn-IOM or IOM used as described in Table 1; "Murch" denotes IOM extracted from Murchison. Results are the average of multiple analyses on a single experimental charge, where the error bars give one standard deviation. Error bars for CO are larger than the measured values, and CO data should be interpreted as a qualitative detection.

Some pressure dependencies for the measured gas compositions were observed. The magnitudes of the trends are summarized in Table 3. Overall, pressure most strongly affects $H_2$ yields at temperatures below 500 °C, with increasing production at increasing pressures and some sample dependency. $CH_4$ and $CO_2$ exhibit minor pressure effects, and yields tend to decrease with increasing pressure at temperatures above 250 °C.

Production of $H_2$ at 250 °C increases by a factor of ~2× from 1 kbar to 3 kbar. At 350 °C, increasing production with increasing pressure is also observed, by as much as a factor of 8. This trend is seen in both the syn-IOM and the Murchison IOM samples and extends up to 10 kbar. At 500 °C, the results from HC096 suggest almost a 4× increase from 3 kbar to 10 kbar, and results from HC113 show a ~3× increase from 3 kbar to 10 kbar. However, IOM HC096 shows a slight decrease in $H_2$ production from 1 kbar to 3 kbar.

CO was infrequently observed, and no conclusions can be drawn about the pressure dependence of its generation at 250 °C or 350 °C. Data from the HC096 sample at 500 °C suggest that CO production may decrease with increasing pressure. However, we note that the CO abundances were very low, and these results should be interpreted with caution.

Table 3. Summary of pressure effects. Comparisons are made only between experiments with the same starting syn-IOM or IOM sample (third column). The absolute change in gas production is given in the fifth column, and the relative change gives the ratio of the gas production from the higher pressure to the lower pressure.

| Gas | Temperature | Sample(s) | Pressure change, kbar | Change in gas production, mmole gas/mg IOM | Relative change |
|---|---|---|---|---|---|
| $H_2$ | 250 °C | HC113 | 1 kbar to 3 kbar | $7.2 \times 10^{-7}$ | 2.3× |
| | 350 °C | HC096 | 1 kbar to 3 kbar | $1.1 \times 10^{-5}$ | 8.1× |
| | | HC083 | 1 kbar to 10 kbar | $2.4 \times 10^{-6}$ | 1.6× |
| | | Murchison | 3 kbar to 10 kbar | $2.0 \times 10^{-5}$ | 4.3× |
| | 500 °C | HC096 | 1 kbar to 3 kbar | $-3.5 \times 10^{-5}$ | 0.2× |
| | | | 3 kbar to 10 kbar | $1.8 \times 10^{-5}$ | 3.6× |
| | | | 1 kbar to 10 kbar | $-1.7 \times 10^{-5}$ | 0.6× |
| | | HC113 | 3 kbar to 10 kbar | $2.0 \times 10^{-5}$ | 2.9× |
| $CH_4$ | 350 °C | HC096 | 1 kbar to 3 kbar | $-3.5 \times 10^{-5}$ | 0.7× |
| | | HC083 | 1 kbar to 3 kbar | $-3.1 \times 10^{-5}$ | 0.8× |
| | | | 3 kbar to 10 kbar | $-4.3 \times 10^{-5}$ | 0.7× |
| | | | 1 kbar to 10 kbar | $-7.4 \times 10^{-5}$ | 0.6× |
| | | Murchison | 3 kbar to 10 kbar | $9.9 \times 10^{-5}$ | 1.5× |
| $CO_2$ | 500 °C | HC096 | 1 kbar to 3 kbar | $-2.3 \times 10^{-4}$ | 0.8× |

For $CH_4$, decreasing abundance with increasing pressure (0.6 to 0.8×) is observed at 350 °C, for both the HC096 and HC083 samples. A very slight decrease from 1 kbar to 3 kbar may be present at 250 °C for HC113, but the difference is so small that in the absence of the 350 °C trend, it would not be notable. No apparent pressure trend is observable for $CH_4$ at 500 °C. Note also that production of $CH_4$ from Murchison IOM shows the opposite trend at 350 °C, with $CH_4$ abundance increasing by 1.5× as pressure increases from 3 kbar to 10 kbar.

The HC096 samples show a slight decrease of 0.8× for $CO_2$ production at 500 °C from 1 kbar to 3 kbar, but no apparent change from 3 kbar to 10 kbar. No other pressure trends are observed for $CO_2$.

While pressure affects gas production with up to nearly an order of magnitude difference for $H_2$, temperature is the dominant variable in controlling the abundance of $CH_4$ and $CO_2$ production. $CH_4$ production increases with increasing temperature, while $CO_2$ displays more complex behavior. The changes in gas production as a function of temperature are summarized in Table 4.

The production of $H_2$ generally increases with increasing temperature by up to a factor of 9.4×, although a slight decrease (0.6×) in $H_2$ is seen from 350 °C to 500 °C for the HC096 sample at 3 kbar. This trend is not repeated for the Murchison sample or the HC113 sample over the same pressure and temperature interval and may suggest some sample-dependence.

The HC096 sample exhibits an apparent increase in CO production with increasing temperature from 250 °C to 500 °C at 3 kbar. However, no CO production was observed for the HC096 sample at 350 °C and 3 kbar, and CO abundances have large uncertainties that make the reported detections more qualitative than quantitative.

The abundance of $CH_4$ increases with increasing temperature. This is best typified by the HC096 sample at 3 kbar from 250 °C to 350 °C (78.2×), and again from 350 °C to 500 °C (37.8×). A more modest increase in $CH_4$ is also observed in the Murchison sample at 3 kbar from 350 °C to 500 °C (8.5×), and in the HC113 sample at 10 kbar from 350 °C to 500 °C (4.6×).

While the abundances of most gases generally increase with increasing temperature, $CO_2$ production is seen to peak at 350 °C in these experiments. Again, this is most easily observed with the HC096 sample at 3 kbar, which shows an increase in $CO_2$ production from 250 °C to 350 °C (11.7×), and then a decrease from 350 °C to 500 °C (0.4×). In contrast, both the Murchison IOM sample at 3 kbar and the HC113 sample at 10 kbar show a very slight increase in $CO_2$ production from 350 °C to 500 °C (1.1×), but this increase is significantly less than the $CH_4$ increase over the same temperature range. The net result is an increase in the $CH_4/CO_2$ ratio in the product gas from 350 °C to 500 °C.

The molar $CH_4/CO_2$ ratios in the experimental products are plotted in Figure 2. For these experiments, $CO_2$ was the dominant product below 500 °C, including both syn-IOM and Murchison IOM. The syn-IOM samples show a decrease in $CH_4/CO_2$ with increasing pressure at 250 °C and 350 °C. The Murchison IOM shows the opposite trend at 350 °C, with increasing pressure resulting in increasing $CH_4/CO_2$. Comparison of Murchison IOM production rates for each gas to syn-IOM HC083 suggests that the increasing $CH_4/CO_2$ results from a slight increase from Murchison IOM relative to the syn-IOM in $CH_4$ production and a slight decrease in $CO_2$ production. That is, whatever structural differences there are between the syn-IOM and the Murchison IOM that controls their hydrous pyrolysis behavior affects the production of both gases. At 500 °C, $CH_4$ abundances become similar to those of $CO_2$, although which gas ultimately dominates the composition is sample-dependent. While HC096 has a higher H/C ratio (0.96) than Murchison (0.70) or HC091 (0.91), the highest H/C ratio is for HC113 (1.13). The $CH_4/CO_2$ ratios of the different samples at 500 °C (HC096>HC113>HC091>Murchison) may be influenced by the H/C ratios of the starting materials (HC113>HC096>HC091>Murchison), but the HC113 and HC096 samples suggest additional factors may also come into play. The production of $CO_2$ is low for HC096, which further suggests that the $CH_4/CO_2$ ratio at 500°C is affected by both O and H in the IOM.

**Table 4.** Summary of the observed temperature effects. Comparisons are made only between experiments with the same starting syn-IOM or IOM sample (third column). The absolute change in gas production is given in the fifth column, and the relative change gives the ratio of the gas production from the higher temperature to the lower temperature.

| Gas | Pressure, kbar | Sample(s) | Temperature change, °C | Change in gas production, mmole gas/mg IOM | Relative change |
|---|---|---|---|---|---|
| $H_2$ | 3 kbar | HC096 | 250 °C to 350 °C | $1.1 \times 10^{-5}$ | 9.4× |
| | | | 350 °C to 500 °C | $-5.2 \times 10^{-6}$ | 0.6× |
| | | | 250 °C to 500 °C | $5.5 \times 10^{-6}$ | 5.4× |
| | | Murchison | 350 °C to 500 °C | $5.4 \times 10^{-5}$ | 5.0× |
| | 10 kbar | HC113 | 350 °C to 500 °C | $1.9 \times 10^{-5}$ | 2.7× |
| $CH_4$ | 3 kbar | HC096 | 250 °C to 350 °C | $6.8 \times 10^{-5}$ | 78.2× |
| | | | 350 °C to 500 °C | $2.5 \times 10^{-3}$ | 37.8× |
| | | | 250 °C to 500 °C | $2.6 \times 10^{-3}$ | 2950.2× |
| | | Murchison | 350 °C to 500 °C | $1.4 \times 10^{-3}$ | 8.5× |
| | 10 kbar | HC113 | 350 °C to 500 °C | $2.7 \times 10^{-3}$ | 4.6× |
| $CO_2$ | 3 kbar | HC096 | 250 °C to 350 °C | $2.0 \times 10^{-3}$ | 11.7× |
| | | | 350 °C to 500 °C | $-1.2 \times 10^{-3}$ | 0.4× |
| | | | 250 °C to 500 °C | $7.6 \times 10^{-4}$ | 5.1× |
| | | Murchison | 350 °C to 500 °C | $3.9 \times 10^{-4}$ | 1.1× |
| | 10 kbar | HC113 | 350 °C to 500 °C | $1.8 \times 10^{-4}$ | 1.1× |

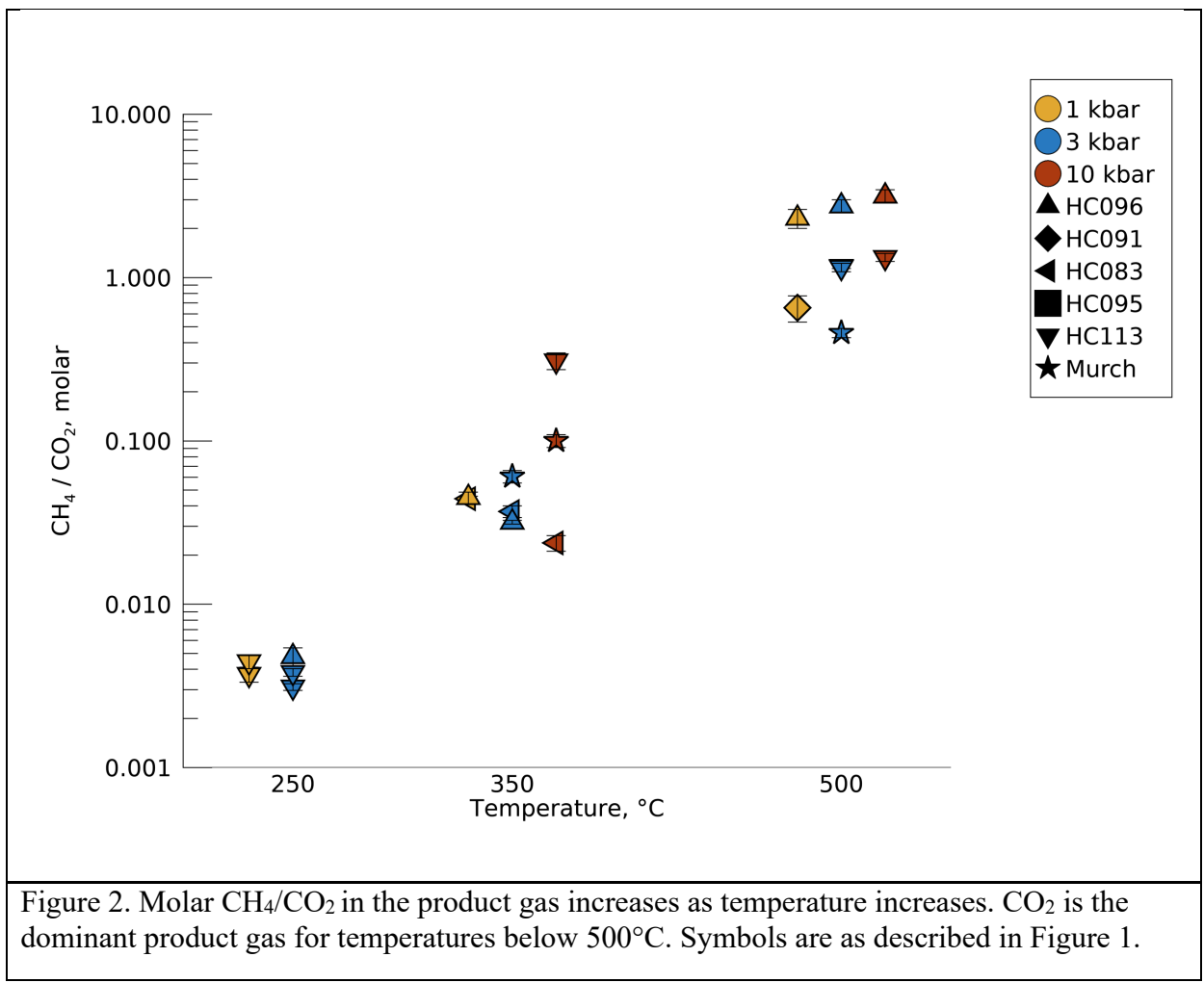

Figure 2. Molar CH$_4$/CO$_2$ in the product gas increases as temperature increases. CO$_2$ is the dominant product gas for temperatures below 500°C. Symbols are as described in Figure 1.

*3.2      Residue composition*

Subsequent to the heating experiments, the solid residues remaining in each experimental charge were analyzed for their bulk elemental (Table 5) compositions. Normalized elemental compositional changes are calculated as:

$$R_{X/C} = \frac{\left(\frac{X}{C}\right)_{final}}{\left(\frac{X}{C}\right)_{initial}}$$

where $X/C$ is the atomic ratio of H, N, or O to C, "initial" refers to the starting materials, and "final" refers to the solids after reaction. The relative changes are graphically displayed in Figure 3, and trends with pressure and temperature are summarized in Table 6.

Table 5. Elemental composition of bulk solid residues after reaction. Changes listed are relative to starting compositions for the syn-IOM and IOM samples listed in Table 1. "Murch." represents IOM extracted from Murchison.

| Experiment | T, °C | P, kbar | (syn)-IOM sample | H/C (at.) | $R_{H/C}$* | N/C (at.) | $R_{N/C}$* | O/C (at.) | $R_{O/C}$* | C, wt.% | H, wt.% | N, wt.% | O, wt.% |
|---|---|---|---|---|---|---|---|---|---|---|---|---|---|
| 250#3 | 250 | 1 | HC095 | 0.73±0.01 | 0.76±0.03 | 0.045±0.000 | 0.79±0.01 | 0.218±0.002 | | 67.4±0.1 | 4.08±0.08 | 3.54±0.01 | 19.6±0.2 |
| 250-1-1-Mar2023 | 250 | 1 | HC113 | 0.75±0.01 | 0.67±0.02 | 0.033±0.001 | 0.33±0.02 | 0.145±0.001 | 0.54±0.02 | 74.3±0.2 | 4.67±0.09 | 2.83±0.09 | 14.4±0.1 |
| 250-1-2-Mar2023 | 250 | 1 | HC113 | 0.78±0.02 | 0.69±0.02 | 0.033±0.003 | 0.33±0.04 | 0.142±0.002 | 0.52±0.02 | 74.1±0.7 | 4.79±0.09 | 2.86±0.26 | 14.0±0.1 |
| 250#2 | 250 | 3 | HC096 | 0.97±0.14 | 1.01±0.15 | 0.052±0.001 | 0.92±0.01 | 0.182±0.025 | | 65.8±0.3 | 4.82±0.09 | 4.03±0.02 | 14.5±0.1 |
| 250-3-1-Mar2023 | 250 | 3 | HC113 | 0.82±0.02 | 0.73±0.03 | 0.034±0.001 | 0.34±0.02 | 0.144±0.003 | 0.53±0.02 | 72.1±1.5 | 4.95±0.10 | 2.85±0.07 | 13.8±0.1 |
| 250-3-2-Mar2023 | 250 | 3 | HC113 | 0.85±0.02 | 0.75±0.03 | 0.031±0.001 | 0.31±0.02 | 0.131±0.002 | 0.49±0.02 | 73.7±0.5 | 5.19±0.10 | 2.70±0.07 | 12.9±0.1 |
| 350_1_1 | 350 | 1 | HC083 | 0.58±0.01 | 0.69±0.04 | 0.052±0.001 | 0.71±0.02 | 0.086±0.002 | 0.23±0.01 | 76.9±1.2 | 3.69±0.07 | 4.64±0.04 | 8.8±0.1 |
| 350_1_2 | 350 | 1 | HC096 | 0.66±0.01 | 0.68±0.04 | 0.034±0.001 | 0.60±0.01 | 0.106±0.001 | | 77.6±0.3 | 4.24±0.08 | 3.07±0.05 | 11.0±0.1 |
| 350_3_1 | 350 | 3 | HC083 | 0.56±0.01 | 0.68±0.04 | 0.050±0.001 | 0.68±0.02 | 0.089±0.001 | 0.23±0.01 | 77.2±0.4 | 3.62±0.07 | 4.47±0.05 | 9.2±0.1 |
| 350_3_2 | 350 | 3 | HC096 | 0.66±0.01 | 0.69±0.04 | 0.058±0.001 | 1.03±0.03 | 0.116±0.003 | | 72.3±1.5 | 3.98±0.08 | 4.93±0.02 | 11.2±0.1 |
| 350-3-M | 350 | 3 | Murch. | 0.63±0.08 | 0.90±0.12 | 0.023±0.001 | 0.63±0.03 | 0.096±0.011 | 0.53±0.06 | 68.2±1.0 | 3.59±0.47 | 1.79±0.06 | 8.7±0.1 |
| 350_10_1 | 350 | 10 | HC083 | 0.87±0.02 | 1.04±0.05 | 0.047±0.002 | 0.64±0.03 | 0.047±0.001 | 0.12±0.00 | 70.6±0.5 | 5.09±0.10 | 3.84±0.15 | 4.4±0.0 |
| 350-10-Mar2023 | 350 | 10 | HC113 | 0.66±0.05 | 0.58±0.05 | 0.022±0.004 | 0.22±0.05 | 0.047±0.003 | 0.18±0.01 | 76.0±5.3 | 4.17±0.08 | 1.97±0.34 | 4.8±0.1 |
| 350-10-M | 350 | 10 | Murch. | 0.64±0.05 | 0.92±0.08 | 0.016±0.001 | 0.44±0.02 | 0.118±0.012 | 0.65±0.07 | 61.2±0.9 | 3.27±0.26 | 1.14±0.03 | 9.6±1.0 |
| 500#1 | 500 | 1 | HC096 | 0.43±0.02 | 0.45±0.03 | 0.055±0.011 | 0.97±0.20 | 0.040±0.001 | | 72.7±2.1 | 2.63±0.05 | 4.68±0.94 | 3.8±0.0 |
| 500#5 | 500 | 1 | HC091 | 0.22±0.00 | 0.24±0.01 | 0.027±0.003 | 0.37±0.01 | 0.028±0.000 | | 81.5±0.1 | 1.50±0.03 | 2.55±0.02 | 3.0±0.0 |
| 500#2 | 500 | 3 | HC096 | 0.40±0.01 | 0.42±0.03 | 0.053±0.007 | 0.93±0.12 | 0.040±0.002 | | 73.1±3.1 | 2.46±0.05 | 4.49±0.53 | 3.9±0.0 |
| 500-3-1-Mar2023 | 500 | 3 | HC113 | 0.36±0.02 | 0.32±0.02 | 0.017±0.002 | 0.17±0.02 | 0.021±0.001 | 0.08±0.01 | 87.6±3.9 | 2.61±0.05 | 1.74±0.21 | 2.5±0.0 |
| 500-3-2-Mar2023 | 500 | 3 | HC113 | 0.59±0.01 | 0.52±0.02 | 0.018±0.000 | 0.18±0.01 | 0.022±0.000 | 0.08±0.00 | 86.9±0.4 | 4.27±0.08 | 1.82±0.01 | 2.5±0.0 |
| 500-3-M | 500 | 3 | Murch. | | | 0.015±0.024 | 0.42±0.07 | | | 55.9±2.1 | | 0.99±0.15 | |
| 500#4 | 500 | 10 | HC096 | 0.46±0.02 | 0.47±0.03 | 0.034±0.003 | 0.60±0.05 | 0.058±0.002 | | 69.6±2.0 | 2.64±0.05 | 2.79±0.23 | 5.4±0.1 |

| 500-10 | 500 | 10 | HC113 | 0.26±0.01 | 0.23±0.01 | 0.012±0.001 | 0.12±0.01 | 0.026±0.000 | 0.09±0.00 | 88.9±0.3 | 1.91±0.04 | 1.23±0.05 | 3.0±0.0 |

*$R_{X/C} = (X/C)_f/(X/C)_i$, where subscripts f and i denote final and initial values and X denotes H, N, or C

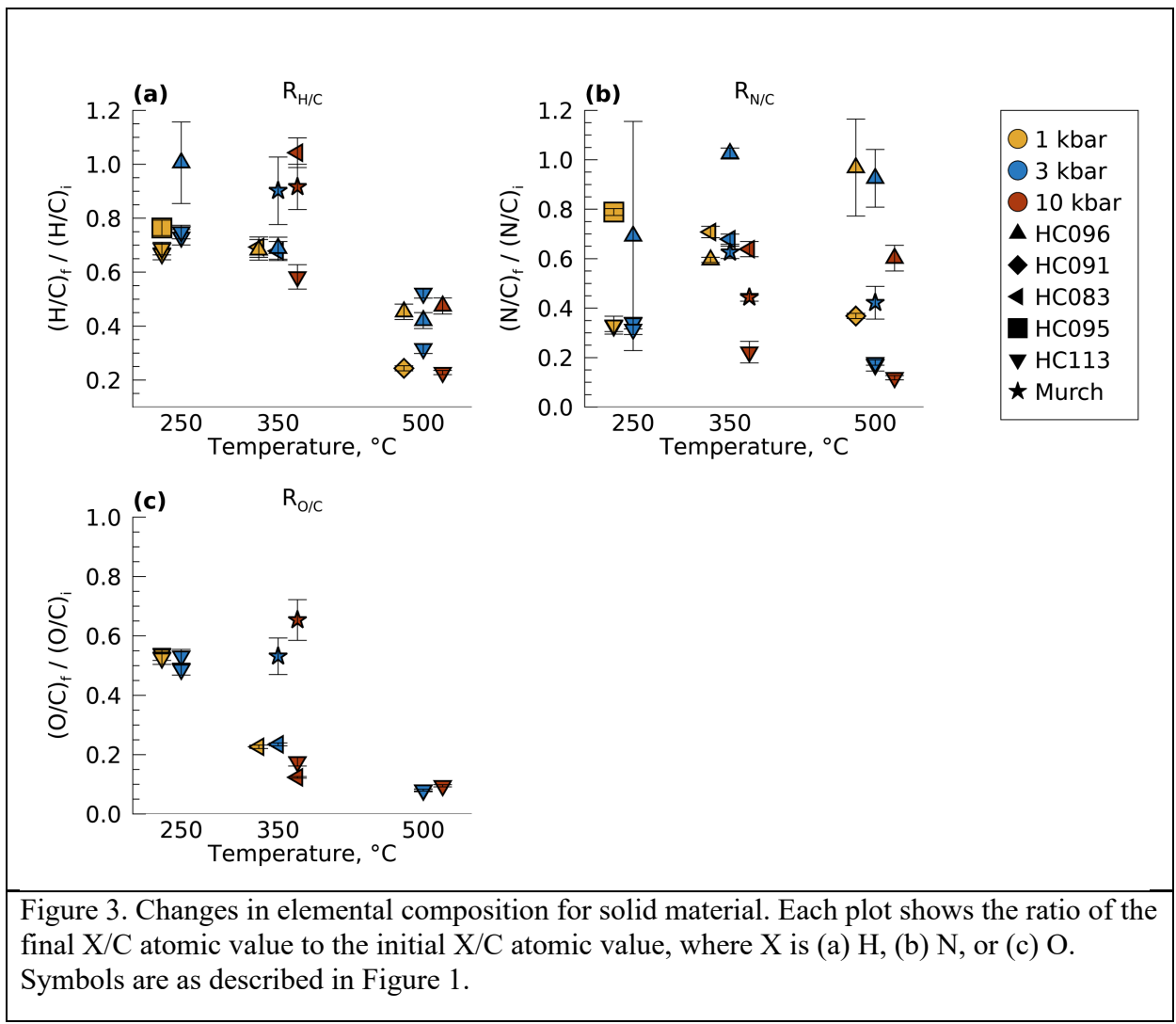

Figure 3. Changes in elemental composition for solid material. Each plot shows the ratio of the final X/C atomic value to the initial X/C atomic value, where X is (a) H, (b) N, or (c) O. Symbols are as described in Figure 1.

The $R_{H/C}$ values generally drop as temperature increases, consistent with the loss of H from the organic matter at a greater rate than the loss of C. The $R_{H/C}$ values at a given temperature may increase very slightly with increasing pressure (e.g., $R_{H/C}$ at 350 °C and 500 °C for HC096 samples), but the averaged values are within one standard deviation. The increase in $R_{H/C}$ with increasing pressure is consistent with the stronger pressure dependence of $H_2$ compared to $CO_2$ or $CH_4$ (Figure 1). The decrease in H/C of complex organic matter following heating is very well established across a broad range of samples (e.g. van Krevelen, 1982). Our analyses overall follow the expected trend of increasingly C-enriched composition at higher temperatures (e.g. Figure 4).

Table 6. Trends in evolution of elemental composition of solids. Values reported are averages of all syn-IOM samples; Murchison IOM is excluded. Cited uncertainties correspond to one standard deviation.

| Temp, °C | Pressure, kbar | $R_{H/C}$ | $R_{N/C}$ | $R_{O/C}$ |
|---|---|---|---|---|
| 250 °C | 1 | 0.71±0.05 | 0.48±0.27 | 0.53±0.01 |
| | 3 | 0.83±0.15 | 0.52±0.34 | 0.51±0.03 |
| 350 °C | 1 | 0.69±0.01 | 0.65±0.08 | 0.23 |
| | 3 | 0.68±0.01 | 0.85±0.25 | 0.23 |
| | 10 | 0.81±0.33 | 0.43±0.29 | 0.15±0.04 |
| 500 °C | 1 | 0.35±0.15 | 0.67±0.42 | |
| | 3 | 0.42±0.10 | 0.43±0.43 | 0.08±0.00 |
| | 10 | 0.35±0.17 | 0.36±0.34 | 0.09 |

The $R_{N/C}$ values display a stronger dependence on the starting material than $R_{H/C}$, even

with the normalization to the initial N/C ratio. The HC113 sample shows the greatest reduction

in N/C, while the HC096 sample shows very little change in N/C. Critically, the $R_{N/C}$ value of the

Murchison sample is intermediate to HC113 and HC096, even though the N loss from

Murchison is substantially lower than either HC113 or HC096. For each given sample, $R_{N/C}$

generally decreases with increasing temperature, although the HC096 sample shows greater loss

of C than N for the 350 °C and 3 kbar experiment. A very slight decrease in $R_{N/C}$ with increasing

pressure is seen, most notably for the HC083 samples at 350 °C, but also for HC096 at 500 °C

and HC113 at 250 °C. This may reflect the slight reduction in production of $CO_2$ and $CH_4$

observed with increasing pressure (Figure 1).

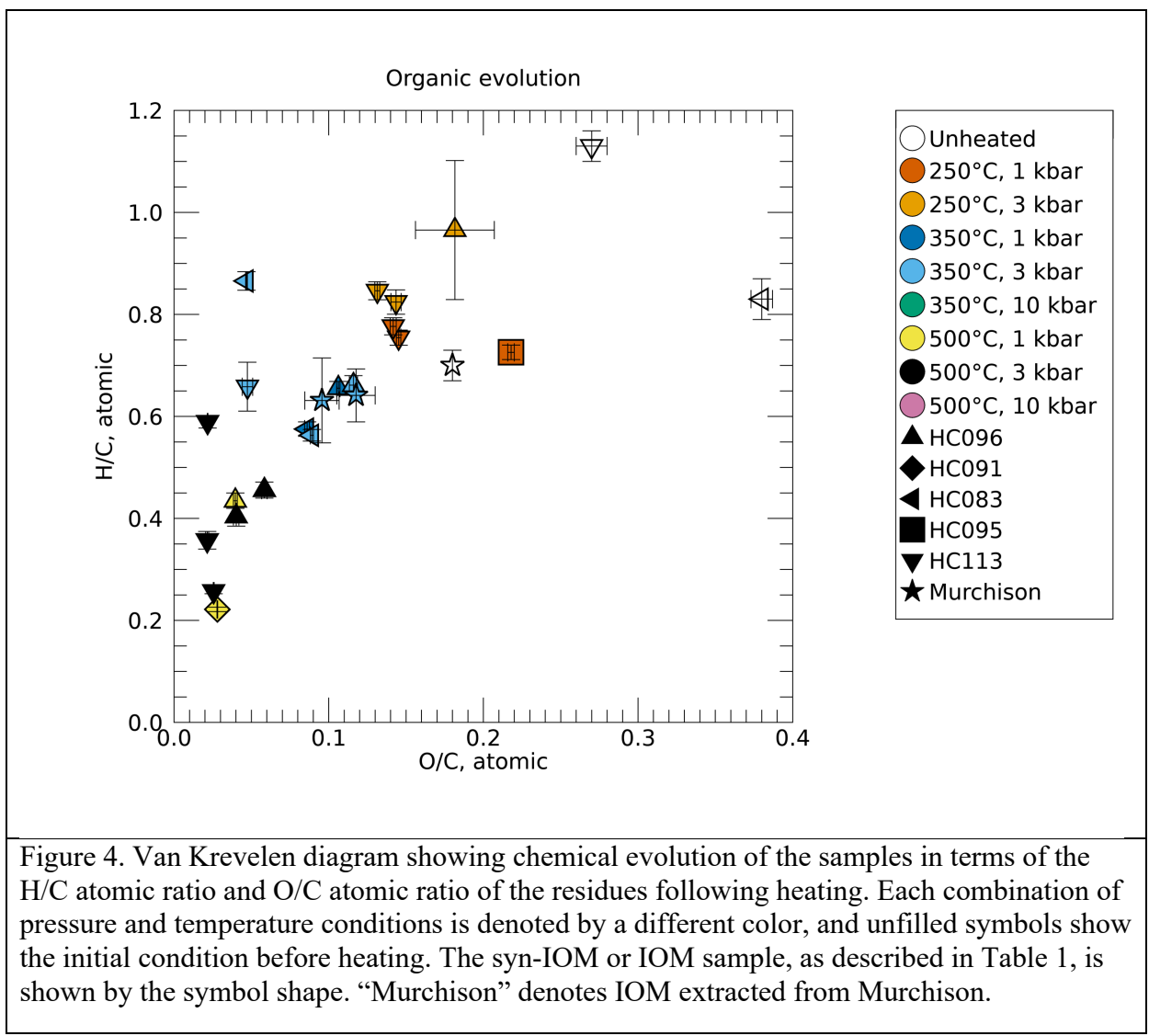

Figure 4. Van Krevelen diagram showing chemical evolution of the samples in terms of the H/C atomic ratio and O/C atomic ratio of the residues following heating. Each combination of pressure and temperature conditions is denoted by a different color, and unfilled symbols show the initial condition before heating. The syn-IOM or IOM sample, as described in Table 1, is shown by the symbol shape. "Murchison" denotes IOM extracted from Murchison.

For the HC113 syn-IOM sample, the $R_{O/C}$ values decrease with increasing temperature, especially from 250 °C to 350 °C. Some leveling off is apparent at 500 °C. Some pressure dependence from 3 kbar to 10 kbar may be evident in the HC083 syn-IOM sample, but no pressure dependence is seen in the HC113 syn-IOM, suggesting that this effect is either marginal or sample-dependent. No significant trend in $R_{O/C}$ is apparent for the Murchison samples. This may be consistent with the observations above (Figure 2) of evidence for structural differences between the syn-IOM and the Murchison IOM that affect production of $CO_2$; however, since there are no data for Murchison IOM at 250°C it is possible that the lack of a trend from 350°C to 500 °C mirrors the leveling off observed in the syn-IOM samples over this temperature range.

The isotopic characterization of the solid residue material is reported in Table 7. Changes are expressed as:

$$\Delta_I = \delta I_{final} - \delta I_{initial}$$

Where $I$ refers to the δD, $\delta^{15}N$, $\delta^{18}O$, and $\delta^{13}C$ values. The relative isotopic changes are displayed in Figure 5, and trends with pressure and temperature are summarized in Table 8.

Table 7. Isotopic composition of bulk solid residues after reaction. Changes listed are relative to starting compositions for the IOM samples listed in Table 1.

| Experiment | T, °C | P, kbar | IOM sample | $\delta^{13}C$, ‰ | $\Delta_{13C}$** | $\delta D$, ‰ | $\Delta_D$** | $\delta^{15}N$, ‰ | $\Delta_{15N}$** | $\delta^{18}O$, ‰ | $\Delta_{18O}$** |
|---|---|---|---|---|---|---|---|---|---|---|---|
| 250#3 | 250 | 1 | HC095 | -11.3±0.1 | -1.1±0.2 | -79.9 | -134.9 | 0.55±0.03 | -0.38±0.17 | 0.8 | |
| 250-1-1-Mar2023 | 250 | 1 | HC113 | -10.3±0.1 | 0.1±0.5 | -88.2 | -51.2 | -2.21±0.16 | -4.61±0.34 | 15.3 | -23.9±0.5 |
| 250-1-2-Mar2023 | 250 | 1 | HC113 | -10.5±0.1 | -0.1±0.5 | -89.1 | -52.1 | -1.80±0.63 | -4.20±0.70 | 8.9 | -30.3±0.5 |
| 250#2 | 250 | 3 | HC096 | -11.2 | -0.8 | -105.0 | -18.0 | 5.68 | 3.68±0.99 | -3.1 | |
| 250-3-1-Mar2023 | 250 | 3 | HC113 | -10.4±0.2 | 0.0±0.5 | -91.6 | -54.6 | -1.80±0.38 | -4.20±0.48 | 12.6 | -26.6±0.5 |
| 250-3-2-Mar2023 | 250 | 3 | HC113 | -10.8±0.1 | -0.4±0.5 | -92.0 | -55.0 | -1.74±0.02 | -4.14±0.30 | 12.0 | -27.2±0.5 |
| 350_1_1 | 350 | 1 | HC083 | -10.0±0.0 | -0.3±0.0 | -97.1 | -129.4 | 0.20±0.15 | -0.36±0.41 | 0.7 | -24.2±0.3 |
| 350_1_2 | 350 | 1 | HC096 | -10.9±0.1 | -0.5±0.1 | -94.1 | -7.1 | 0.89±0.17 | -0.41±0.17 | 0.4 | |
| 350_3_1 | 350 | 3 | HC083 | -9.9±0.0 | -0.2±0.0 | -94.1 | -126.4 | -0.01±0.07 | -0.56±0.39 | -11.6 | -36.5±0.3 |
| 350_3_2 | 350 | 3 | HC096 | -10.8±0.0 | -0.4±0.0 | -62.5 | 24.5 | 1.66±0.52 | 0.36±0.52 | -4.0 | |
| 350-3-M | 350 | 3 | Murchison | -18.9±0.3 | -0.5±0.3 | -27.4±6.4* | -822.4±8.8* | 14.5±2.5 | 10.1±2.5 | -10.1±3.9 | -49.3±3.9 |
| 350_10_1 | 350 | 10 | HC083 | -10.3±0.5 | -0.6±0.5 | -87.6 | -119.9 | -0.73±0.07 | -1.28±0.39 | -8.8 | -33.7±0.3 |
| 350-10-Mar2023 | 350 | 10 | HC113 | -11.9±2.1 | -1.5±2.2 | -83.8 | -46.8 | -1.80±0.35 | -4.20±0.46 | -31.0 | -70.2±0.5 |
| 350-10-M | 350 | 10 | Murchison | -19.9±1.4 | -1.5±1.4 | 28.0±34.3 | -767.0±34.8 | -2.15±1.09 | -6.50±1.10 | -14.9±2.8 | -54.1±2.8 |
| 500#1 | 500 | 1 | HC096 | -11.2±0.1 | -0.8±0.1 | -95.3 | -8.3 | 5.73±0.98 | 4.43±0.98 | -36.0 | |
| 500#3 | 500 | 1 | HC091 | -11.6±0.0 | -1.5±0.0 | -101.8 | -137.8 | 1.71±0.03 | 1.11±0.08 | -34.1 | |
| 500#2 | 500 | 3 | HC096 | -11.4±0.1 | -1.0±0.1 | -84.1 | 2.9 | 3.73±0.39 | 2.43±0.39 | -25.5 | |
| 500-3-1-Mar2023 | 500 | 3 | HC113 | -10.8±1.1 | -0.38±1.2 | -82.4 | -45.4 | -1.67±0.19 | -4.07±0.36 | -25.6 | -64.8±0.5 |
| 500-3-2-Mar2023 | 500 | 3 | HC113 | -10.2±0.1 | 0.2±0.5 | -90.5 | -53.5 | -1.54±0.35 | -3.94±0.46 | -35.7 | -74.9±0.5 |
| 500-3-M | 500 | 3 | Murchison | -20.7±1.8 | -2.3±1.8 | -79.0±2.0 | -874.0±6.3 | -1.07±1.76 | -5.42±1.76 | -5.3±1.0 | -44.5±1.1 |
| 500#4 | 500 | 10 | HC096 | -12.3±0.0 | -1.9±0.0 | -100.2 | -13.2 | 7.93±0.70 | 6.63±0.70 | -21.9 | |
| 500-10 | 500 | 10 | HC113 | -11.1±0.1 | -0.7±0.5 | -103.1 | -66.1 | -2.33±0.15 | -4.73±0.33 | -15.1 | -54.3±0.5 |

*Rubber contamination observed in vial; δD may be affected

**$\Delta_I = I_f - I_i$, where subscripts f and i denote final and initial values and I denotes $\delta^{13}C$, $\delta D$, $\delta^{15}N$, or $\delta^{18}O$.

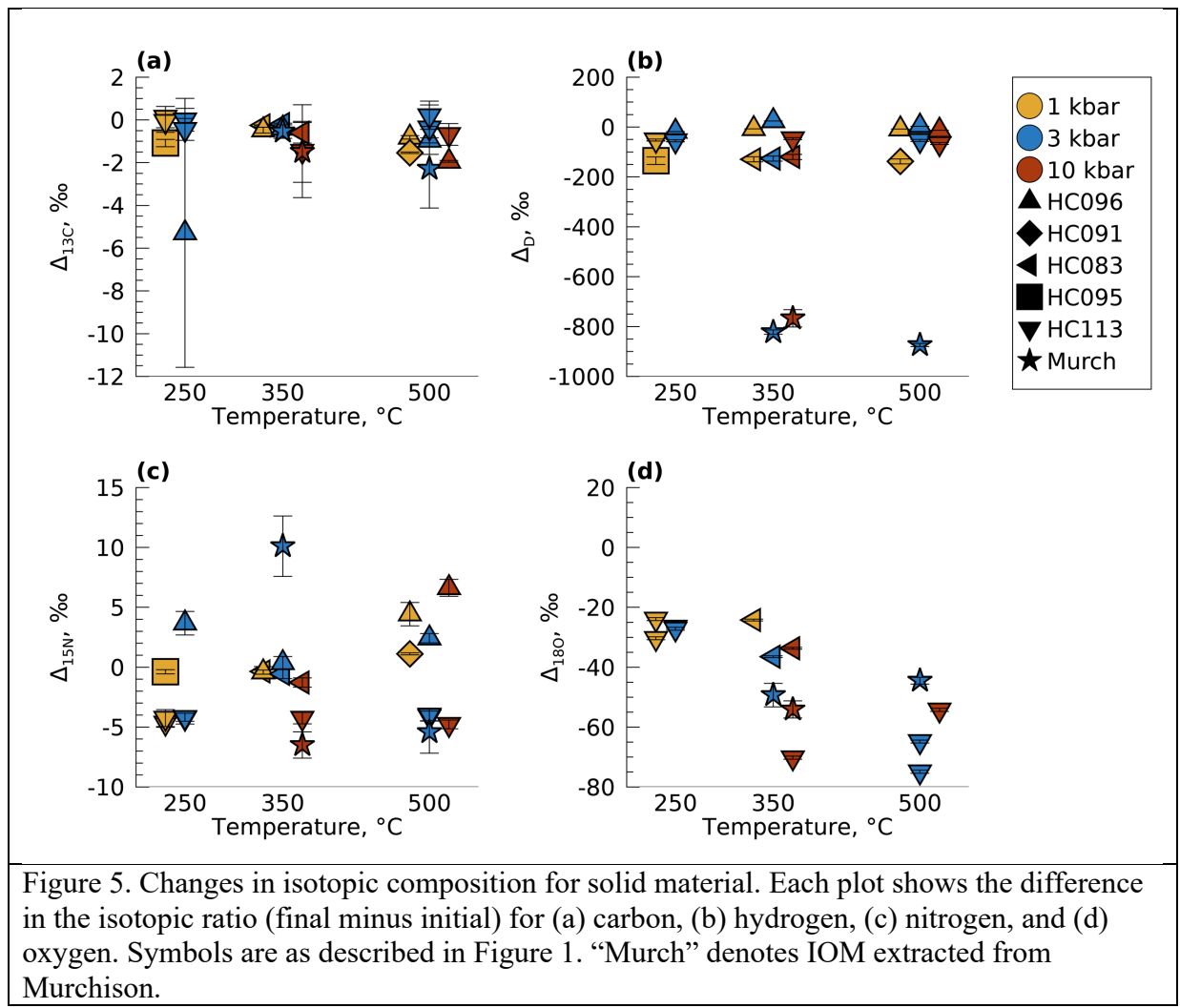

Figure 5. Changes in isotopic composition for solid material. Each plot shows the difference in the isotopic ratio (final minus initial) for (a) carbon, (b) hydrogen, (c) nitrogen, and (d) oxygen. Symbols are as described in Figure 1. "Murch" denotes IOM extracted from Murchison.

In general, the $\delta^{13}C$ value for the final residue is lower than for the starting materials, suggesting that $^{13}C$ may be preferentially lost during reaction. However, in most experiments the $\Delta_{13C}$ value is within error of zero, and changes in $\delta^{13}C$ are overall small. With increasing temperature and pressures of 1 kbar or 10 kbar, the loss of $^{13}C$ may increase slightly; however, no trend with temperature is apparent in the 3 kbar data. $CO_2$ production is stable from 350 °C to 500 °C at 3 kbar but decreases from 350 °C to 500 °C at 1 kbar and 10 kbar (Figure 1). This may suggest that the loss of $^{13}C$ is more strongly tied to production of $CO_2$ rather than $CH_4$. This is consistent with previous studies of IOM, which found that $\delta^{13}C$ is higher in carbonyl groups that may yield $CO_2$ during decomposition, as compared to aliphatic and aromatic carbon (Oba and Naraoka, 2009). The behavior of $\Delta_{13C}$ appears to be similar between the syn-IOM and the Murchison IOM samples.

Evolution of $\delta D$ is more dramatic than $\delta^{13}C$, consistent with previous studies at ambient pressure (Oba and Naraoka, 2009), but also displays a sample dependence, especially for Murchison IOM. Notably, no sample dependence was observed in the $R_{H/C}$ values, which may point to multiple different H reservoirs in the starting material with different isotopic characteristics. While values for $\Delta_D$ suggest the preferential volatilization of D, no trends with pressure or temperature are apparent. The $\Delta_{18O}$ values also suggest preferential volatilization of $^{18}O$ relative to $^{16}O$, with $\Delta_{18O}$ becoming more negative with increasing temperature for 3 kbar and 10 kbar.

While the C, H, and O isotopic systems all display a counter-intuitive tendency to preferentially volatilize the heavier isotopes from the IOM, leading to more depleted residues, the evolution of the N isotopic system is more variable. The 1 kbar data show an increase in $\Delta_{15N}$ as temperature increases, as do the 10 kbar data. Scatter amongst data points at 3 kbar make trends difficult to identify. The HC113 samples at 3 kbar show equally depleted residues at both 250 °C and 500 °C. This is fairly consistent with the HC096 data points at 3 kbar, and this sample displays greater depletion at 350 °C relative to either 250 °C or 500 °C. Examination of the HC083 data points at 350 °C and the HC096 data points at 500 °C suggests a slightly increased depletion in the heavy isotopes of the residues with increasing pressure. While the Murchison samples at 350 °C and 10 kbar, and 500 °C at 3 kbar are generally consistent with the syn-IOM samples, the Murchison 350 °C and 3 kbar experiment shows preferential volatilization of $^{14}N$, to a much greater extent than any of the other experiments.

Table 8. Trends in the evolution of the isotopic compositions of solids. Values reported are averages of all syn-IOM samples. Murchison IOM is excluded from these averages since its behavior differs; see Table 7 for individual values. Cited uncertainties correspond to one standard deviation.

| Temp, °C | Pressure, kbar | $\Delta_{13C}$ | $\Delta_D$ | $\Delta_{15N}$ | $\Delta_{18O}$ |
|---|---|---|---|---|---|
| 250 °C | 1 | -0.34±0.65 | -79.4±48.1 | -3.06±2.33 | -27.1±4.5 |
| | 3 | -0.43±0.42 | -42.5±21.2 | -1.55±4.53 | -26.9±0.4 |
| 350 °C | 1 | -0.37±0.16 | -68.2±86.5 | -0.39±0.04 | -24.2 |
| | 3 | -0.21±0.18 | -50.9±106.7 | 5.09±9.00 | -40.4±5.57 |
| | 10 | -1.02±0.43 | -349.4±462.3 | -2.21±1.72 | -52.9±18.4 |
| 500 °C | 1 | -1.18±0.51 | -73.1±91.6 | 2.77±2.35 | |
| | 3 | -0.73±0.85 | -242.5±421.8 | -1.41±3.16 | -62.6±13.6 |
| | 10 | -1.31±0.89 | -39.7±37.4 | 0.95±8.03 | -54.3 |

### 3.3 $NH_4^+$

Data related to $\Sigma NH_3$ in the final solutions are presented in Table 9. The $\Sigma NH_3$ abundances are representative of $NH_3$ and $NH_4^+$, due to the acidification of the solution during sample preparation. Hereafter, we refer to the two interchangeably. Results are summarized in Figure 6. In general, a significant percentage of the N present in the starting material was extracted in the form of $NH_3$, and the efficiency of extraction is noticeably different between the syn-IOM and the Murchison IOM, with less N extracted from Murchison. Extraction of N increases from 250 °C to 500 °C. In the syn-IOM, no additional extraction is observed from 350 °C to 500 °C, potentially because very little N remains in the solid form by 350 °C. For the Murchison IOM, a more modest increase in N extraction is seen from 350 °C to 500 °C. The total abundance of extracted N is on the order of μmoles per mg of IOM. There is some evidence for isotopic depletion of the $NH_3$ with increasing pressure, but these effects are slight for the syn-IOM. $NH_3$ from the Murchison IOM is much more enriched compared to the starting material at 3 kbar pressures, and the enrichment is more notable at 350°C than at 500°C.

Table 9. Abundance and isotopic characterization of product $NH_4^+$. $\Delta_{15N,NH4}$ is reported relative to the isotopic composition of the starting solid material. b.d. is below the detection limits.

| Experiment | T, °C | P, kbar | IOM sample | N in $NH_4^+$, µmoles/mg sample | $\delta^{15}N$ of $NH_4^+$, ‰** | $\Delta_{15N,NH4}$* | % of IOM nitrogen in $NH_4^+$ |
|---|---|---|---|---|---|---|---|
| 250-3 | 250 | 1 | HC095 | 1.68 | | | 60.73 |
| 250-1-1 | 250 | 1 | HC113 | 2.60 | 3.23±0.40 | 0.83±0.49 | 56.89 |
| 250-1-2 | 250 | 1 | HC113 | 2.95 | 2.55±0.32 | 0.15±0.41 | 64.44 |
| 250-3-1 | 250 | 3 | HC113 | 2.85 | 2.31±0.29 | -0.09±0.38 | 62.33 |
| 250-3-2 | 250 | 3 | HC113 | 2.74 | 2.34±0.29 | -0.06±0.38 | 59.96 |
| 350-10 | 350 | 10 | HC113 | 4.19 | 5.77±0.72 | 3.37±0.81 | 91.74 |
| 500#1 | 500 | 1 | HC096 | 2.54 | | | 79.96 |
| 500#3 | 500 | 1 | HC091 | 2.09 | | | 90.21 |
| 500-3-1 | 500 | 3 | HC113 | 3.76 | 3.09±0.39 | 0.69±0.48 | 82.25 |
| 500-3-2 | 500 | 3 | HC113 | 4.22 | 1.07±0.13 | -1.33±0.22 | 92.30 |
| 500-10 | 500 | 10 | HC113 | 4.10 | 0.90±0.11 | -1.50±0.20 | 89.62 |
| 350-3-M | 350 | 3 | Murchison | 0.57 | 105±28 | 101±28 | 32.06 |
| 350-10-M | 350 | 10 | Murchison | 0.54 | b.d. | | 30.59 |
| 500-3-M | 500 | 3 | Murchison | 0.72 | 68.2±7.3 | 64±7 | 40.83 |
| Syn-IOM average | 250 | | | 2.56±0.51 | 2.61±0.43 | 0.21 | 60.87±2.81% |
| | 350 | | | 4.19 | 5.77 | 3.36 | 91.74% |
| | 500 | | | 3.34±0.97 | 1.69±1.22 | -0.71 | 86.87±5.41% |

*$NH_4^+$ $\delta^{15}N$ minus the $\delta^{15}N$ value for the bulk starting IOM
**errors estimated as same %error for other $\delta^{15}N$ measurements

The significant difference in the degree of $\Sigma NH_3$ extraction from the syn-IOM and Murchison residues indicates potential differences in the speciation of N in the organic solids. Previous studies have suggested the predominant presence of amines in the molecular structure of chondritic IOM (Alexander et al. 2017); however, the presence of N-bearing heterocycle moieties has also been hypothesized (Derenne and Robert, 2010; Remusat et al., 2005; Vollmer et al., 2024). It is highly likely, therefore, that the N-functional groups in the Murchison IOM are significantly more attached to aromatic structures, resulting in an enhanced refractory behavior relative to the N in the syn-IOM.

The high $\delta^{15}N$ compositions of dissolved $NH_3$ from Murchison IOM in the reactant solutions (68-105 ‰) are in close agreement with previous results ($\delta^{15}N$ = 39-53 ‰) involving hydrothermal alteration of Murchison IOM at 250 °C to 450 °C but lower pressure conditions (500 bars; Foustoukos et al., 2021) as well as at 300°C and 1 kbar with reported $\delta^{15}N$ of 18 ‰ (Pizzarello and Williams, 2012). Similar $\delta^{15}N$ values ($\delta^{15}N$ = 69-70 ‰) have also been measured in $NH_4^+$ extracted via hot water (110 °C, 24 h; Pizzarello et al., 1994) and MgO-distillation at ambient temperatures (Kung and Clayton, 1978) of bulk Murchison samples. The ammonia released from the bulk samples is associated with $NH_4^+$-bearing salts and/or with highly labile N-bearing IOM species (Pizzarello et al., 1994; Pizzarello and Williams, 2012). It seems, therefore, that the N-H bearing functional groups released during hydrothermal treatment of IOM at elevated temperatures could be directly linked to the sources of inorganic N in the bulk meteorite with a potential $\delta^{15}N$ signature of 50 ‰ to 100 ‰. These values are consistent with the estimated $\delta^{15}N$ composition of the free $NH_3$ in the bulk samples returned from the Ryugu asteroid (~ 56 ‰; Hashizume et al., 2024). Recent remote-sensing studies have suggested that comets and carbonaceous bodies from the outer Solar System have large inventories of solid nitrogen compounds, such as ammonium salts (Altwegg et al., 2020; De Sanctis et al., 2015; Poch et al., 2020). Future studies on samples returned from asteroids (e.g., Bennu, Ryugu) would be instrumental in establishing these $\delta^{15}N$ values as signatures of inorganic N in chondritic parent bodies. This could have profound implications for understanding the source and evolution of N in chondritic material, and especially for constraining N cycling between pre- and proto-solar material (Pizzarello and Williams, 2012).

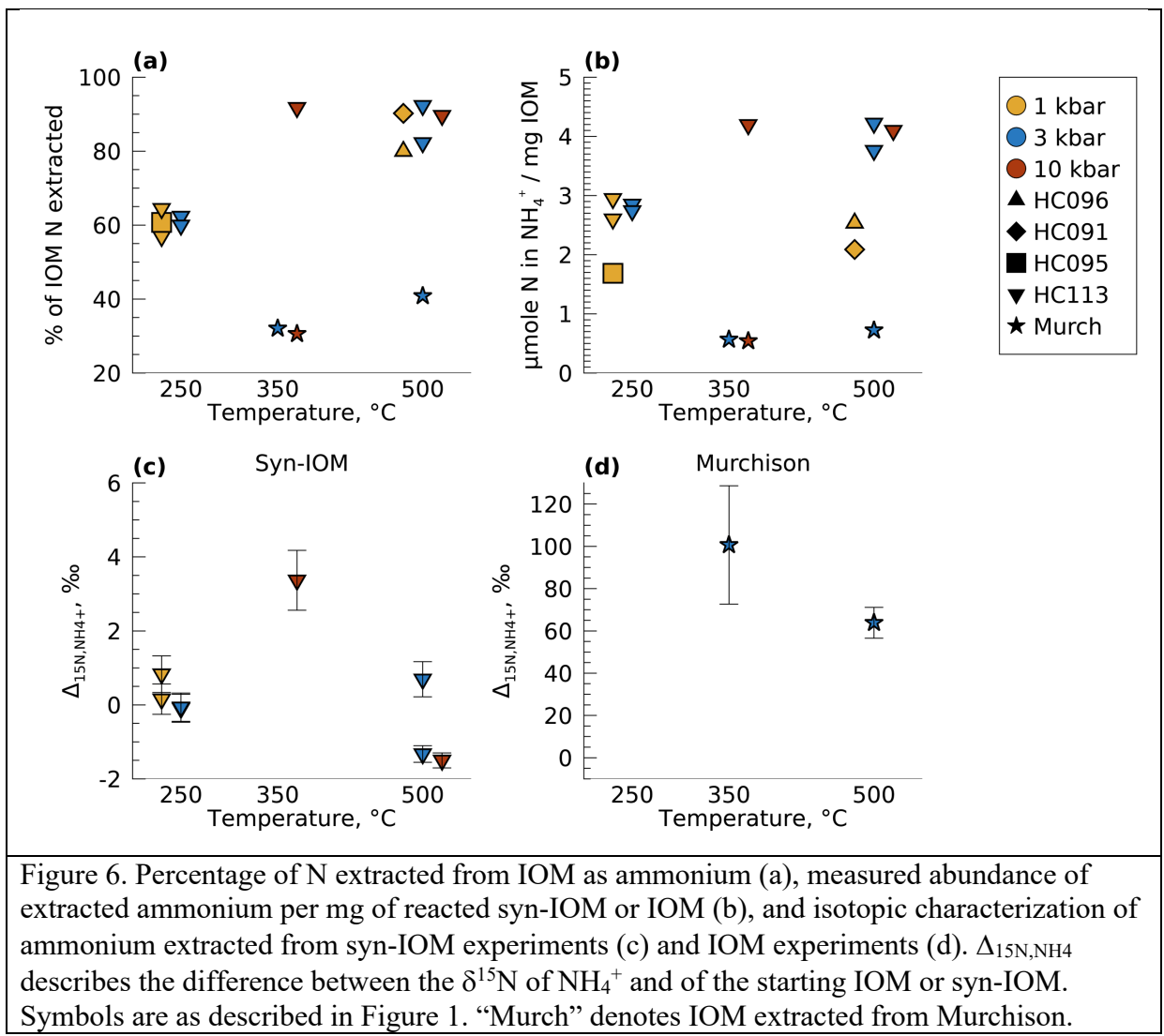

Figure 6. Percentage of N extracted from IOM as ammonium (a), measured abundance of extracted ammonium per mg of reacted syn-IOM or IOM (b), and isotopic characterization of ammonium extracted from syn-IOM experiments (c) and IOM experiments (d). $\Delta_{15N,NH4}$ describes the difference between the $\delta^{15}N$ of $NH_4^+$ and of the starting IOM or syn-IOM. Symbols are as described in Figure 1. "Murch" denotes IOM extracted from Murchison.

## 4 Discussion

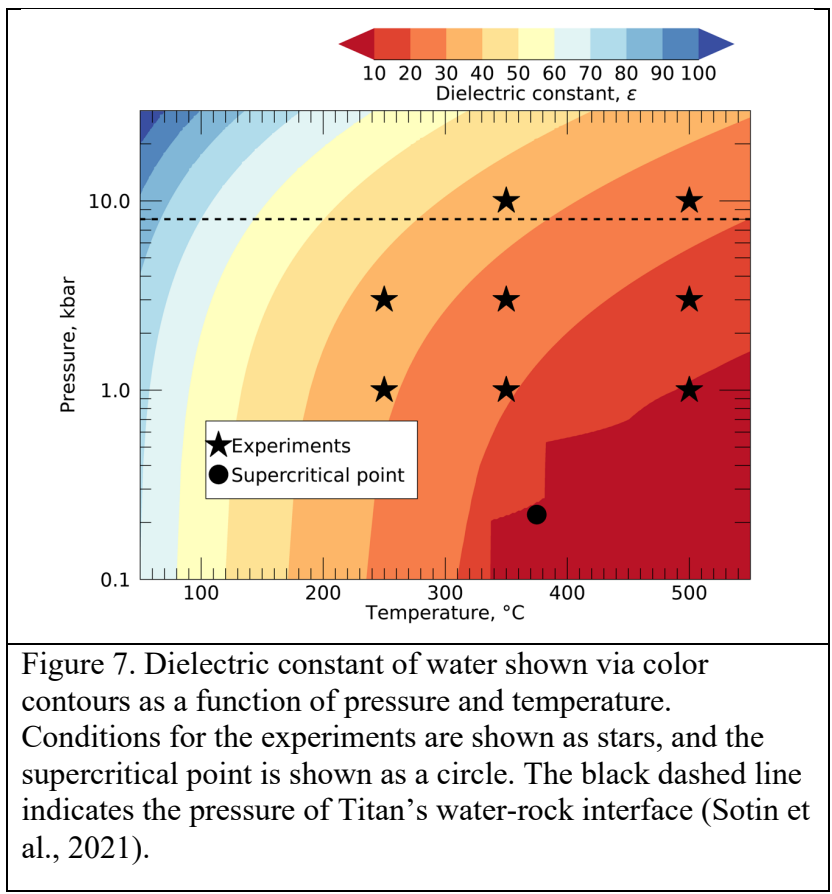

### 4.1 Impact of the dielectric constant and kinetics on gas production

Figure 7. Dielectric constant of water shown via color contours as a function of pressure and temperature. Conditions for the experiments are shown as stars, and the supercritical point is shown as a circle. The black dashed line indicates the pressure of Titan's water-rock interface (Sotin et al., 2021).

The abundance trends observed for different gases as a function of pressure and temperature may be related in part to changes in the dielectric constant, $\varepsilon$, of water (Figure 7). The dielectric constant was calculated using the Deep Earth Water (DEW) model (Sverjensky et al., 2014). Briefly, this model provides an analytical formulation of $\varepsilon$ at high pressures via extrapolation of the observed linear relationship between $\ln(\varepsilon)$ and $\ln(\rho)$, where $\rho$ is the density of water. The formula fits experimental data ranging up to 12 kbar and 600 °C (Fernandez et al., 1997; Heger et al., 1980), and provides estimates for pressure and temperature conditions where experimental data are lacking, especially at higher pressures. Higher values for $\varepsilon$ correspond to increased polarity. For a given temperature, as the pressure increases $\varepsilon$ also increases. At lower temperatures, the change in $\varepsilon$ with pressure is smaller. As temperature increases, the dependence of $\varepsilon$ on pressure also increases. At 350 °C where an increase in $CO_2$ is observed with increasing pressure, the higher polarity of water may lead to more effective solvation and removal of C-O or C=O groups in the Murchison IOM and syn-IOM structures. Pyrolysis experiments on coal samples suggest that product $CO_2$ may derive from such functional groups rather than the incorporation of O from water (Cramer, 2004).

Gas production as a function of $\varepsilon$ is shown in Figure 8. A clear trend of decreasing $H_2$ production with increasing $\varepsilon$, and increasing polarity, is visible. This may indicate that $H_2$ production is associated with nonpolar functional groups in the organic solids. The trend with $\varepsilon$ becomes flatter at higher pressures, where the $\varepsilon$ contours begin to flatten in Figure 7. CO production may also trend with $\varepsilon$, although caution is warranted in using our determinations of CO abundances (see Section 3.1). The behavior of $CH_4$ is similar to that of $H_2$, which may be expected given the relatively nonpolar nature of alkyl groups compared to C-O or C=O groups.

The behavior of $CO_2$ relative to $\varepsilon$ is more complex and suggests that multiple factors may be at play. $CO_2$ initially increases with increasing $\varepsilon$, which may correspond to enhanced solvation of the more polar C-O or C=O groups with increasing polarity of the solvent. However, at $\varepsilon$~25, the $CO_2$ abundance stops increasing. In Figure 7, this corresponds to the salmon-colored contour from approximately 235 °C to 310 °C at 0.1 kbar. This may reflect depletion of extractable C-O at $\varepsilon \sim 25$ or could reflect more complex chemical effects. A similar reduction in the abundances of $CO_2$ at higher temperatures and increases in the $CH_4/CO_2$ ratios have been reported for anthracite coal in anhydrous experiments, with $CO_2$ peaking at 400 °C at 5 kbar and the peak extending to 600 °C at 2 kbar (Mastalerz et al., 1993), suggesting that at higher temperatures $CO_2$ production from complex organic matter may decrease independent of the presence of water. Future characterization of the functionality of experimental residues compared to starting materials may help resolve which answer is more likely.

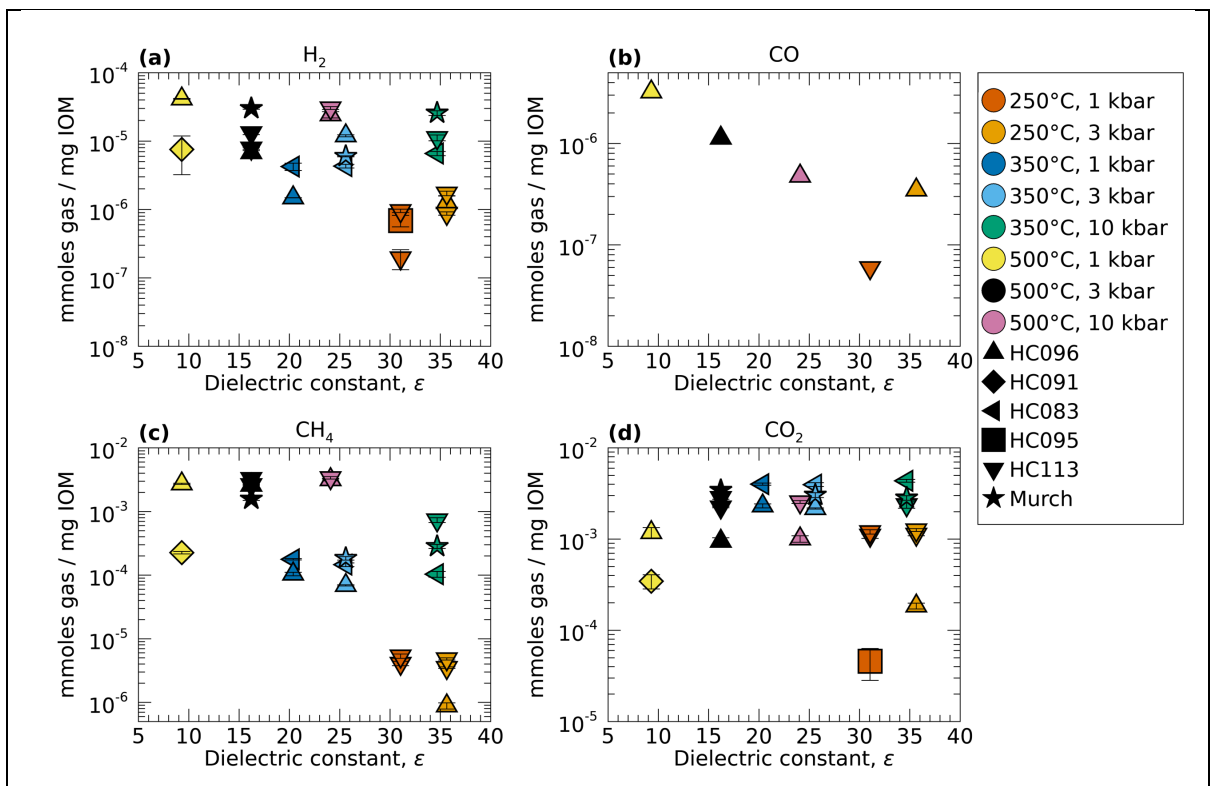

Figure 8. Gas production as a function of dielectric constant for (a) $H_2$, (b) CO, (c) $CH_4$, and (d) $CO_2$. Symbols are as described in Figure 4. "Murch" denotes IOM extracted from Murchison.

Consistent with the complex behavior of gas production, kinetic modeling of the O/C and H/C ratios of residues from terrestrial kerogen samples suggests a range of activation energies for precursor functional groups that yield $H_2O$, $CO_2$, $CH_n$, and $CH_4$ gases during thermal maturation on geologic timescales (Burnham and Sweeney, 1989; Burnham 2017). This canonical "Vitrimat" model indicates lower activation energies for $CO_2$ production, and higher energies for $CH_4$ production. Figure 9a shows the O/C and H/C ratios of experimental residues from this work overlaid with variations of the Vitrimat model of Burnham and Sweeney (1989). The original model is shown by the line labeled as "BS89 model." The color of the line shows progressive evolution as temperature increases from 20°C to greater than 600°C at a rate of 10°C/Ma, taken from Burnham and Sweeney (1989). The HC083 model and Murch model utilize identical parameters, except for the starting compositions that have been changed to match the HC083 syn-IOM and Murchison IOM, respectively. The shapes of these curves reasonably reproduces the corresponding datapoints, shown by the left-pointing triangle (HC083) and star (Murchison) symbols. The color of the symbols corresponds to the experimental temperatures; data at all pressures are shown but not differentiated, since pressure is the weaker variable. While the overall evolutionary trend is a reasonable match for the data, the experimental temperatures are noticeably higher than the model temperatures. The strongest model parameter controlling the temperature is the activation energy for the gas precursors, and this difference may indicate that the model underestimates the activation energies, which are distributed here from 38 to 74 kcal/mol. Burnham and Sweeney (1989) note the same mismatch in their comparison to laboratory data, with higher temperatures required in laboratory settings compared to geologic settings, and cite pressure as one important variable, especially for materials with higher starting O/C. A second version of this figure with the same distribution of activation energies uniformly shifted 22 kcal/mol higher is shown in Figure 9b to demonstrate the effect of activation energy on the modeled temperature. The HC113 model in both panels is further tuned to better fit data from the HC113 experiments (downward-pointing triangles); here, the distribution of activation energies for the $CO_2$ precursors are shifted higher with respect to the $H_2O$, $CH_n$, and $CH_4$ precursors to change the shape of the curve (Table 10).

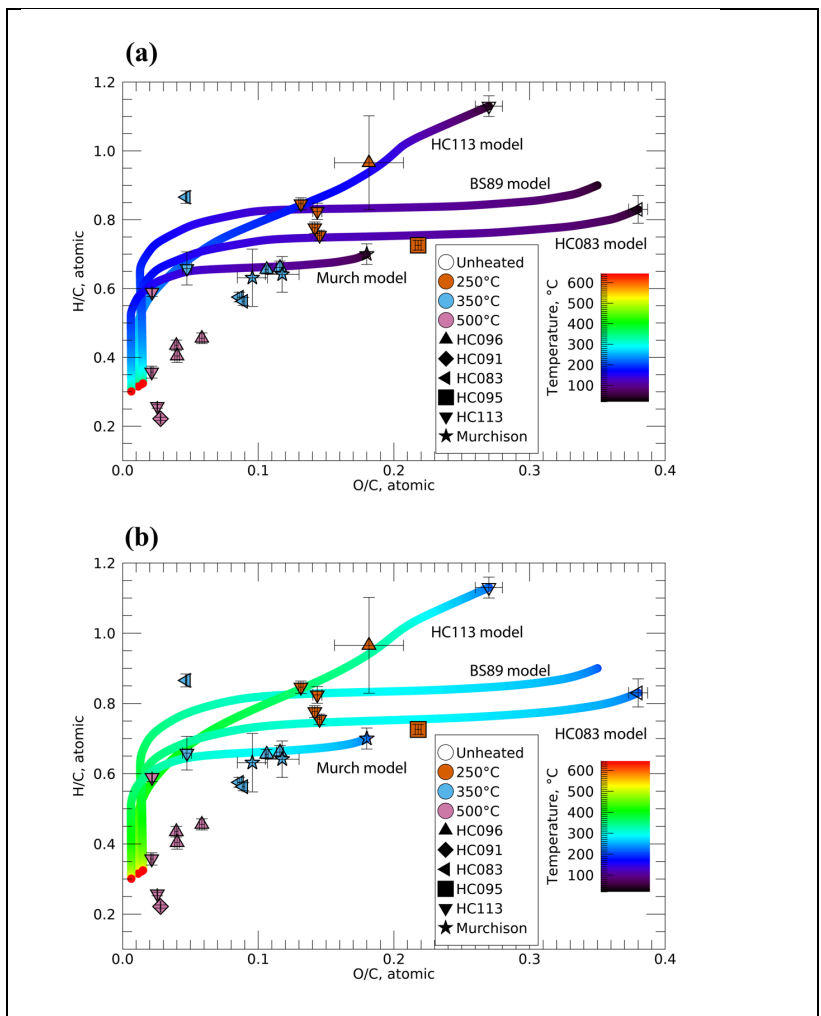

**Figure 9.** Kinetic models of the O/C and H/C ratio of solid residues are shown as lines, and the changing color corresponds to temperature as shown by the colorbar. (a) Models are based on the original Vitrimat model of Burnham and Sweeney (1989), shown by the "BS89 model" line. Data from residues generated in this work are overlaid as points, with symbol shapes corresponding to the starting syn-IOM or IOM material. The symbol color shows the temperature for the corresponding experiment. Data include all pressures. The HC083 model and Murch model utilize identical kinetic parameters to the BS89 model, except for the O/C and H/C of the starting material which reflects the corresponding syn-IOM (HC083) or IOM (Murchison) sample. The HC113 model has been additionally tuned to better match the data, with higher activation energies for $CO_2$ precursors as summarized in Table 10. (b) Same as panel (a), but the activation energy distributions are all shifted 22 kcal/mol higher.

**Table 10.** Activation energies used for Vitrimat modeling of $CO_2$ in the original model (middle column) and for HC113 in this work (right column).

| Activation energy E in Figure 9a, kcal/mol | Percent of $CO_2$ precursors, (Burnham and Sweeney, 1989) | Percent of $CO_2$ precursors, this work |
|---|---|---|
| 42 | 5 | |
| 44 | 15 | |
| 46 | 25 | |
| 48 | 25 | |
| 50 | 15 | 5 |
| 52 | 10 | 15 |
| 54 | 5 | 25 |
| 56 | | 25 |
| 58 | | 15 |
| 60 | | 10 |
| 62 | | 5 |

This may suggest that $CO_2$ precursors in the HC113 sample differ from the other samples. This is consistent with the different productivity of $CO_2$ at 250°C from this sample compared to the others (Figure 1); however, generally speaking, the model is under-constrained and additional work is needed to confirm the nature of these potential differences.

### 4.2  Equilibrium speciation in Titan's interior

The predominance fields were calculated for aqueous species in the COH system between 1 kbar and 10 kbar for the COH system using the DEW model (Sverjensky et al., 2014), which has been validated for pressures up to 60 kbar. The calculations utilized values from Zhang and Duan (2005) to calculate the density of water. For pressures above 5 kbar, Sverjensky et al. (2014) was used to calculate the dielectric constant, while the SUPCRT equation formulated in Johnson and Norton (1991) was used for lower pressures.

With increasing temperature, the stability field for $CO_2$ increases (Figure 10), while that of $CH_4$ decreases. The pH at which $CO_2$ transitions to $HCO_3^-$ increases relative to neutral pH with increasing temperature. The transition from $CH_4$ to $HCO_3^-$ and $CO_3^{2-}$ occurs at conditions that are more oxidizing with increasing temperature. The transition from $CO_2$ to $CH_4$ is relatively insensitive to pressure (Figure 10a), while speciation between $HCO_3^-$ and $CO_3^{2-}$ ionic species and $CH_4$ varies somewhat with increasing pressure. As pressure increases, the range over which $CH_4$ dominates decreases. $CH_4$ stability under equilibrium conditions is favored overall by lower temperatures and lower pressures. The equilibrium speciation of aqueous compounds with one C in Figure 11 provides another visualization of this trend. Figure 11a shows the speciation at 8 kbar, the pressure at Titan's water-rock interface, while Figure 11b shows the speciation at 2 kbar, a pressure that is equivalent to several 10s of km below Titan's icy crust and within the ocean layer (Sotin et al., 2021). For outgassing from the rocky interior, fluid parcels would travel from 8 kbar through 2 kbar. At low temperatures, C speciation drives towards $CH_4$ as the stable phase, especially for the lower pressures. Here, we assume slightly oxidizing conditions (fayalite-magnetite-quartz buffer, $f_{O2}$ = FMQ, +1) and basic conditions (3 units above neutral pH).

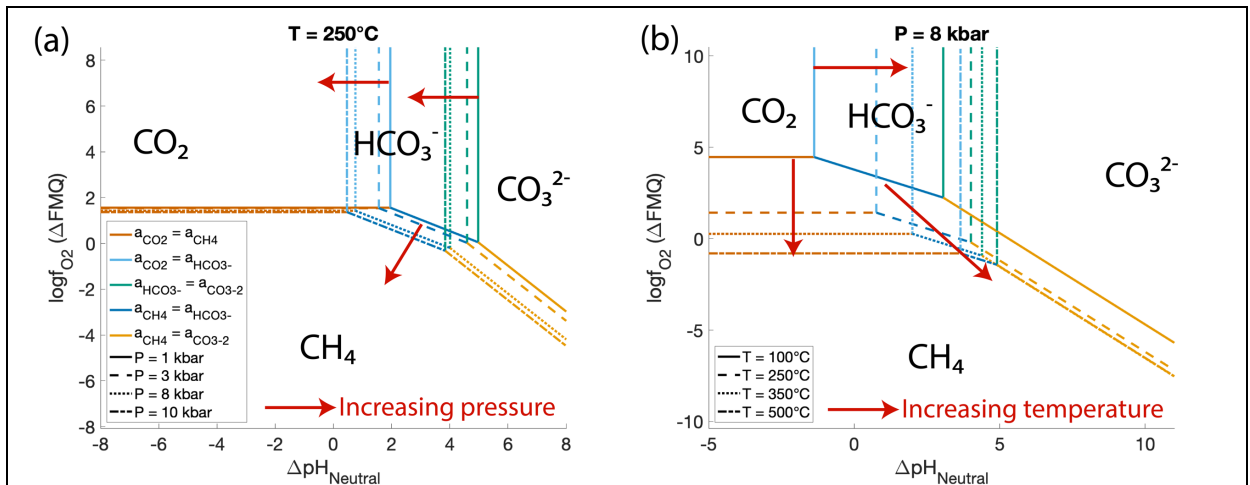

Figure 10. Predominance plots for major compounds in the COH system. Regions in redox-pH space where each compound is most abundant at equilibrium are marked, and the lines show the redox-pH condition at which the chemical activities of the two compounds on either side of the line are equal. The identity of the two compounds is given by the line color. (a) Trends as pressure increases at a constant temperature of 250°C. Pressure increases from 1 kbar (solid line) to 3 kbar (dashed line) to 8 kbar (dotted line) to 10 kbar (dash dot line). The red arrow shows the direction that each boundary in the predominance plot moves with increasing pressure. Panel (b) holds pressure constant at 8 kbar, and shows the effect of increasing temperature: 100 °C (solid line), 250 °C (dashed line), 350°C (dotted line), and 500 °C (dash dot line). Line colors in panel (b) are the same as in panel (a).

Given the relatively short (~ 48 hour) timescale that these experiments were run, C-O-H equilibrium was not achieved except for experiments at 500 °C and 10 kbar. Our experimental results indicate that $CH_4$ is favored by higher temperatures. At 3 kbar for the HC096 sample, the molar ratio of $CH_4/CO_2$ increases from $5\times10^{-3}$ at 250 °C to 2.7 at 500 °C. A previous hydrothermal experiment on IOM from the Murray CM2 chondrite identified the loss after heating of carbonyl groups in solid residues via NMR (Yabuta et al., 2007). It is possible that decarboxylation of IOM side chains is the dominant source of volatile C, and at higher temperatures, the kinetics and extent of $CO_2$ reduction to $CH_4$ is higher (Foustoukos, 2012). In this case, the results do not reflect equilibrium speciation, but constrain overall production of volatile C. The ultimate fate of volatile C depends on the specific environmental context in which it is produced and equilibrated. Volatilization itself is an instantaneous process, geologically speaking, and the quantities produced are well within the solubility limits for relevant pressures and temperatures. Therefore, the volatiles are likely to be entrained in the fluid phase, and to mobilize (or not) as the fluid does. While trapped fluids may eventually reach equilibrium conditions, kinetics are likely to play the more important role in the absence of confining phases.

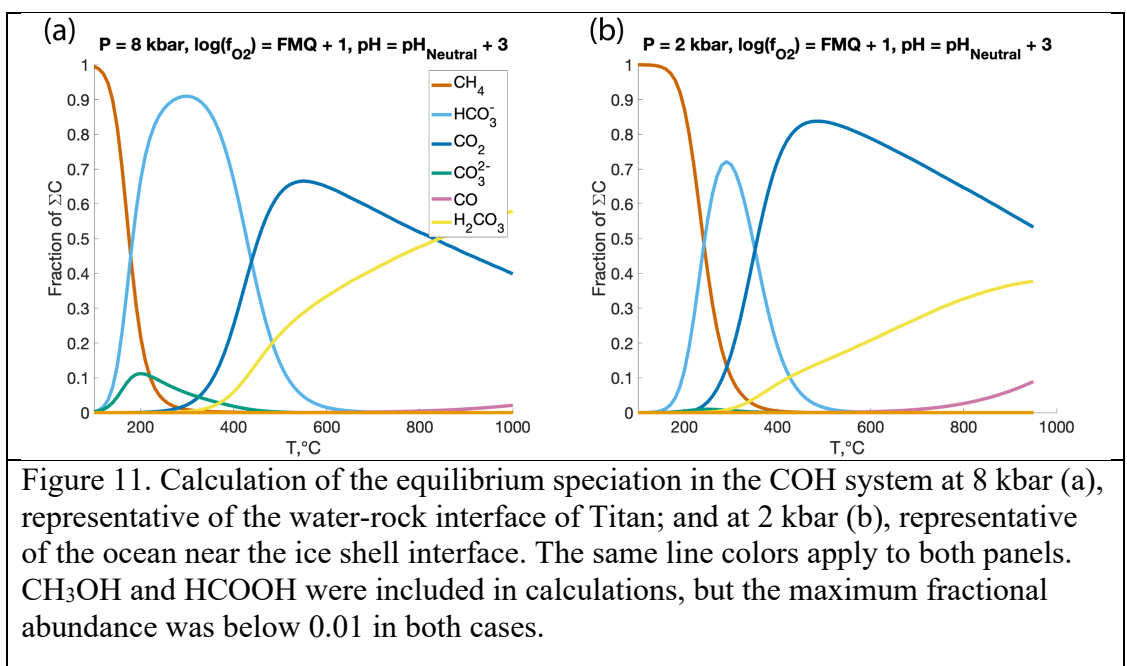

Figure 11. Calculation of the equilibrium speciation in the COH system at 8 kbar (a), representative of the water-rock interface of Titan; and at 2 kbar (b), representative of the ocean near the ice shell interface. The same line colors apply to both panels. $CH_3OH$ and $HCOOH$ were included in calculations, but the maximum fractional abundance was below 0.01 in both cases.

CO was infrequently observed in the hydrous pyrolysis experiments conducted here. This is consistent with equilibrium calculations, which indicate that CO should not be the dominant equilibrium species. Comparisons of measured C speciation in hydrothermal vent fluids to equilibrium models for fluid cooling indicate that equilibration in the C-O-H system is very rapid for temperatures in excess of ~200 °C (Foustoukos et al., 2009). Since CO does not begin to appear in equilibrium speciation plots until approximately 700 °C at $f_{O2}$ = FMQ-2 and 1 pH unit above neutral, and this region of stability is relatively insensitive to pH and $f_{O2}$, CO is unlikely to be a major hydrothermal product of IOM in Titan's interior so long as IOM reaction occurs in the presence of water.

### 4.3  Implications for Titan's atmosphere

At present, Titan's atmosphere includes approximately $9{\times}10^{18}$ kg of $N_2$ and on the order of $2{\times}10^{17}$ kg of $CH_4$ and organic matter that may be photolytically derived from $CH_4$ (Tobie et al., 2012). These values are likely lower limits since they do not consider atmospheric escape over the lifetime of Titan's atmosphere. In order for IOM in Titan's interior to significantly contribute to atmospheric $N_2$ or $CH_4$, production of these components in the interior must be comparable to or in excess of their total atmospheric abundances. Furthermore, outgassing from Titan's interior is likely to be incomplete, with efficiencies on the order of 1 to 6% cited on the basis of $^{40}$Ar abundances in Titan's atmosphere (Miller et al., 2019; Waite et al., 2005). Here, we scale our results to values that are reasonable for Titan's interior and consider whether the isotopic behavior observed in our experiments is consistent with existing constraints.

In Table 11, we summarize a calculation of the total $CH_4$ and $CO_2$ production that might be expected from Titan's interior, in comparison to an existing estimate of the surface C reservoir (Tobie et al., 2012) of approximately $1.3{\times}10^{19}$ moles of C. We use calculations of the total mass of Titan's rocky core with a convecting or a conducting upper ice shell to set the range of interior masses (Sotin et al., 2021). Based on density data, Reynard and Sotin (2023) suggest that the present-day interior of Titan may contain between 17 wt.% and 30 wt.% organic matter, which we use as additional constraints. Note that our premise assumes mass loss from interior organics to generate Titan's atmosphere; this suggests that the present-day value for the interior organic mass is lower than Titan's accreted organic mass. Since temperature is generally the strongest control on production of $CH_4$ and $CO_2$ in our experiments, we use the minimum and maximum abundances for each temperature bin for all pressures and samples, including both Murchison IOM and syn-IOM. Our estimates show that at 250 °C, $CH_4$ production from all organic matter in Titan's interior may be of order 1× to 10× the abundance of the surface reservoir. Since the surface reservoir is a lower limit on the total amount of material, these production rates are barely sufficient or insufficient. A few scenarios are possible as a result.

Table 11. Summary of volatile carbon production in the interior compared to surface carbon reservoir. Values for $CH_4$ and $CO_2$ production show the minimum and maximum for temperature sets that include both syn-IOM and Murchison IOM.

| Mass of Titan's core, $10^{22}$ kg[a] | Wt.% IOM in core[b] | Mass of IOM in Titan's core, $10^{22}$ kg | $CH_4$ production, moles | | | $CO_2$ production, moles | | | Surface reservoir, moles[c] |
|---|---|---|---|---|---|---|---|---|---|
| | | | 250 °C | 350 °C | 500 °C | 250 °C | 350 °C | 500 °C | |
| 7.97 - 9.71 | 17 - 30 | $1.4 - 2.9$ | $1.2 \times 10^{19}$ – $1.6 \times 10^{20}$ | $9.3 \times 10^{20}$ – $2.1 \times 10^{22}$ | $3.1 \times 10^{21}$ – $9.9 \times 10^{22}$ | $6.2 \times 10^{20}$ – $3.7 \times 10^{22}$ | $2.9 \times 10^{22}$ – $1.3 \times 10^{23}$ | $4.7 \times 10^{21}$ – $1.0 \times 10^{23}$ | $1.3 \times 10^{19}$ |

[a]Sotin et al. (2021)
[b]Reynard and Sotin (2023)
[c]Tobie et al. (2012)

Some or most of Titan's $CH_4$ may be derived from accreted ices. Methane ice and/or methane clathrate has been detected remotely at more than ten different comets (Bockelée-Morvan and Biver, 2017) and may be a common component of cometary ices. However, rapid photolytic processing of $CH_4$ at Titan's surface (Yung et al., 1984) suggests an interior source may be important. Alternatively, Titan's atmospheric $CH_4$ might be reconciled with our experimental data at 250°C if the outgassing efficiency from Titan's interior is closer to 10 % to 100 %, at least for $CH_4$. However, this seems unlikely, given the low abundance of Kr in Titan's atmosphere, the potential role of Kr-rich Phase Q in Titan's interior chemistry (Miller et al., 2019), and the assumed similarity between Kr and $CH_4$ outgassing rates (Glein, 2015). Another alternative is that some $CO_2$ produced in the interior may have reacted to yield $CH_4$ over geologically long timescales, as discussed above. Our estimate for the total interior reservoir of volatile C at 250 °C is $6.3 \times 10^{20}$ to $3.7 \times 10^{22}$ moles, or 50 to 2000 times the surface reservoir. If the original IOM abundance in the interior was higher than 30 wt.%, this may be reconciled with present-day constraints for the organic mass in the interior given mass loss via volatilization. Volatilization losses are likely to be on the order of 10s of percent, and such a mechanism does not provide a much larger margin for derivation of the surface reservoir. Finally, higher temperatures may be more representative of the interior environment in which IOM produced $CH_4$. At 350 °C, the minimum production of $CH_4$ is nearly 100 times greater than at 250 °C, and 1 % outgassing efficiency is consistent with the estimated surface reservoir.

Very little O is observed at Titan's surface (Hörst et al., 2008) and outgassing of volatilized $CO_2$ from Titan's interior would contravene the observed surface chemistry. However, $CO_2$ that does not equilibrate in Titan's interior to $CH_4$ may be sequestered as precipitated carbonate minerals. Aragonite ($CaCO_3$) is a common high-pressure carbonate phase that may precipitate from the carbonate system in the presence of $Ca^{2+}$ in aqueous solution in Titan's ocean. As a rough estimate for the activity of $Ca^{2+}$ ($a_{Ca}$), we utilize a value of $10^{-2}$ drawn from the range from $>10^{-2}$ to $>10^{-1}$ calculated for Europa (Melwani Daswani et al., 2021). They utilize a hydrosphere mass for Europa that is only ~40% of Titan's hydrosphere mass and $a_{Ca}$ may be diluted at Titan compared to Europa. However, $Mg^{2+}$ and $Fe^{2+}$ are also common carbonate-forming cations, and $Mg^{2+}$ may be more abundant than $Ca^{2+}$ (Melwani Daswani et al., 2021); the use of $a_{Ca}$ from Europa therefore provides a reasonable estimate for Titan's ocean. We utilize a thermodynamic database from PyGeochemCalc (Awolayo and Tutolo, 2022) for compounds at 5 kbar, and calculate the thermodynamic stability of the carbonate system at 25 °C as a function of pH (Figure 12) with Geochemist's Workbench (Bethke 2022). For pH > 6.7, aragonite is thermodynamically stable when $\log(a_{CO2})$ exceeds -1.4; this corresponds to an $a_{CO2}$ value of approximately 0.04. From Table 11, the minimum predicted abundance of $CO_2$ is $6.2 \times 10^{20}$ moles. Titan's ocean may have a present-day mass of $(1.92-4.04) \times 10^{22}$ kg (Sotin et al., 2021). This suggests a molality on the order of 0.02-0.03 mol/kg. The minimum predicted $CO_2$ abundance is therefore very close to the stability limit for aragonite, and $CO_2$ in excess of this amount may form carbonate precipitates that would sink to Titan's ocean floor. Alternatively, $CO_2$ clathrates may form (Safi et al., 2017) and sink or float depending on the density of Titan's ocean (Idini and Nimmo, 2024; Leitner and Lunine, 2019; Mitri et al., 2014). Like $CH_4$ clathrate, the presence of such a layer may have implications for the thermal evolution of Titan's interior and for surface-interior exchange (Kalousová and Sotin, 2020), though thermal properties of clathrates may vary as a function of composition (Sloan and Koh, 2007).

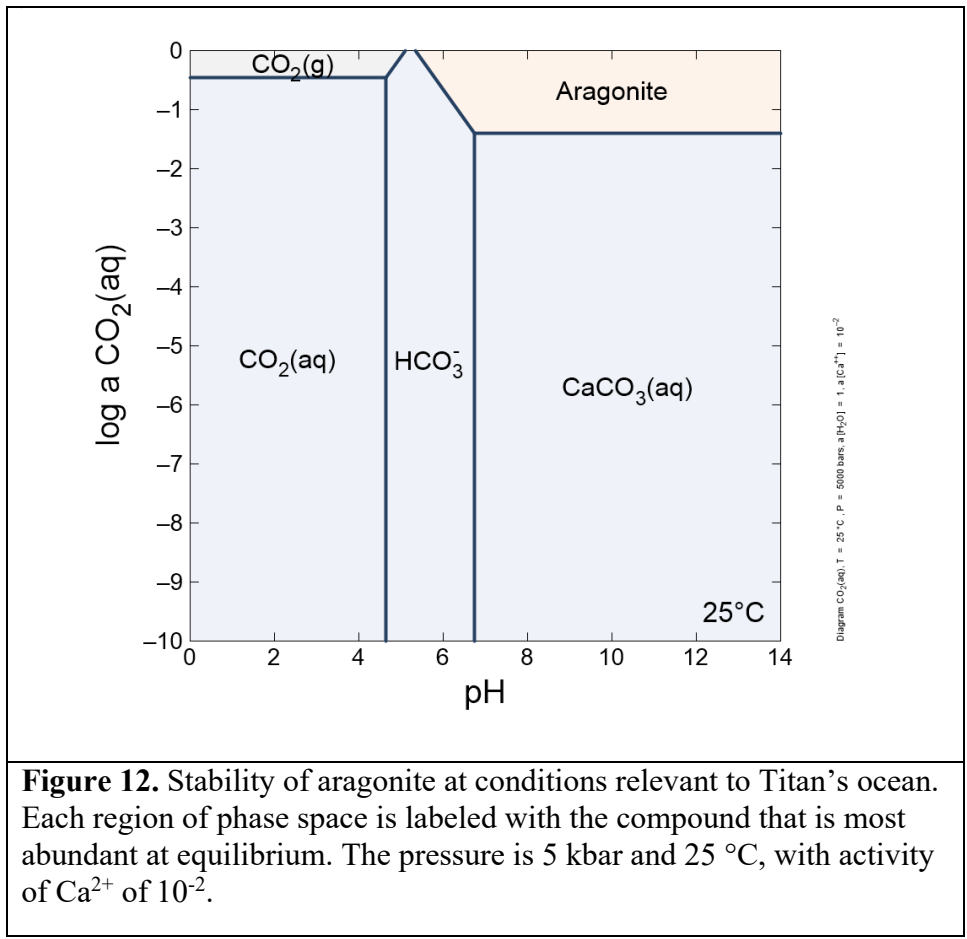

**Figure 12.** Stability of aragonite at conditions relevant to Titan's ocean. Each region of phase space is labeled with the compound that is most abundant at equilibrium. The pressure is 5 kbar and 25 °C, with activity of $Ca^{2+}$ of $10^{-2}$.

Our results suggest that there is little isotopic fractionation between IOM and the labile C species derived from it. The Huygens probe measured $\delta^{13}C$ for Titan's atmospheric $CH_4$ between -8 ‰ to -38 ‰ (Niemann et al., 2010), and this value may have evolved from lower initial atmospheric values (Nixon et al., 2012). This range of values brackets the $\delta^{13}C$ of the Murchison IOM sample used here (-18.5 ‰), suggesting that our experimental results for C isotopic evolution are consistent with IOM as a major source for atmospheric $CH_4$, at least if Murchison IOM is isotopically representative. More broadly, the $\delta^{13}C$ of IOM in chondrites, including Ryugu, generally ranges from -5 ‰ to -35 ‰ (Alexander et al., 2007; Yabuta et al., 2023). CI chondrites and Ryugu may represent primitive samples of the solar system outside of Saturn's orbit (Hopp et al., 2022), and we therefore conclude that this range for $\delta^{13}C$ is a reasonable representation for IOM accreted by Titan.

Table 12. Summary of volatile N production in the interior compared to surface nitrogen reservoir. Values for $NH_3$ production are for Murchison IOM.

| Mass of Titan's core, $10^{22}$ kg[a] | Wt.% IOM in core[b] | Mass of IOM in Titan's core, $10^{22}$ kg | $NH_3$ production, moles | | Surface reservoir, moles of N[c] |
|---|---|---|---|---|---|
| | | | 350 °C | 500 °C | |
| 7.97 - 9.71 | 17 - 30 | 1.4 – 2.9 | $7.3\times10^{21}$ – $1.7\times10^{22}$ | $9.8\times10^{21}$ – $2.1\times10^{22}$ | $6.4\times10^{20}$ |

[a]Sotin et al. (2021)
[b]Reynard and Sotin (2023)

[c]Tobie et al. (2012)

Production of N in Titan's interior may result in a volatile N reservoir that is comparable in abundance to the atmospheric reservoir, if 350 °C is a representative temperature (Table 12). This reservoir may, therefore, be abundant enough to reasonably contribute to Titan's atmospheric N (Miller et al., 2019) if interior processes can transform $NH_3$ to $N_2$ (Glein, 2015), or if $NH_3$ outgasses efficiently enough and surface processes alter it to $N_2$ (Atreya et al., 1978).

Any experimentally produced $N_2$ would be indistinguishable from contaminant air in our GC measurements. However, we can use the abundance of air as an upper limit on $N_2$ production for comparison to the $NH_3$ abundances. Our Murchison experiment "500-3-M" has an air abundance on the order of $1 \times 10^{-4}$ mmole/mg IOM, or 0.1 µmole/mg IOM, comparable to the $NH_3$ measured for that sample (0.72 µmole/mg IOM). However, in general, the abundance of $NH_3$ exceeds the upper limit for $N_2$ by closer to a factor of 100, suggesting that $NH_3$ is the dominant form of volatilized N in our experiments.

Our isotopic results suggest that $NH_3$ derived from chondritic IOM is significantly enriched in $^{15}N$. However, this enrichment may have a relatively small effect on the allowable contribution of N from Titan's interior IOM. Figure 13a shows the nominal mixing model for organic-derived N and $NH_3$ from Miller et al. (2019), which assumes that there is no isotopic fractionation between extracted N and bulk IOM. This model compares reported values for the N-isotopic composition of $N_2$ and $NH_3$ ices as well as IOM to Titan's atmosphere as one constraint on atmospheric origins. The $^{36}Ar/^{14}N$ ratio measured in Titan's atmosphere provides an additional constraint. The ratio for $N_2$ is taken from cometary ices based on the observed correlation of $^{36}Ar$ and $N_2$ (Balsiger et al., 2015). $NH_3$ ice is assumed to have a near-zero $^{36}Ar/^{14}N$ because of temperature fractionation (Mousis et al., 2009). The ratio for organics is linked to the Phase Q noble gas component that is associated with carbonaceous phases in chondrites (Ott et al., 1981; Busemann et al. 2000). Titan's atmosphere intersects the mixing line at roughly 50 % organics and 50 % $NH_3$ ice. In Figure 13b, the model is shown using isotopic values consistent with the maximum observed fractionation. Here, Titan's atmosphere intersects the mixing line at closer to 60 % $NH_3$ and 40% organics. The difference between these two scenarios is close to negligible with current uncertainties.

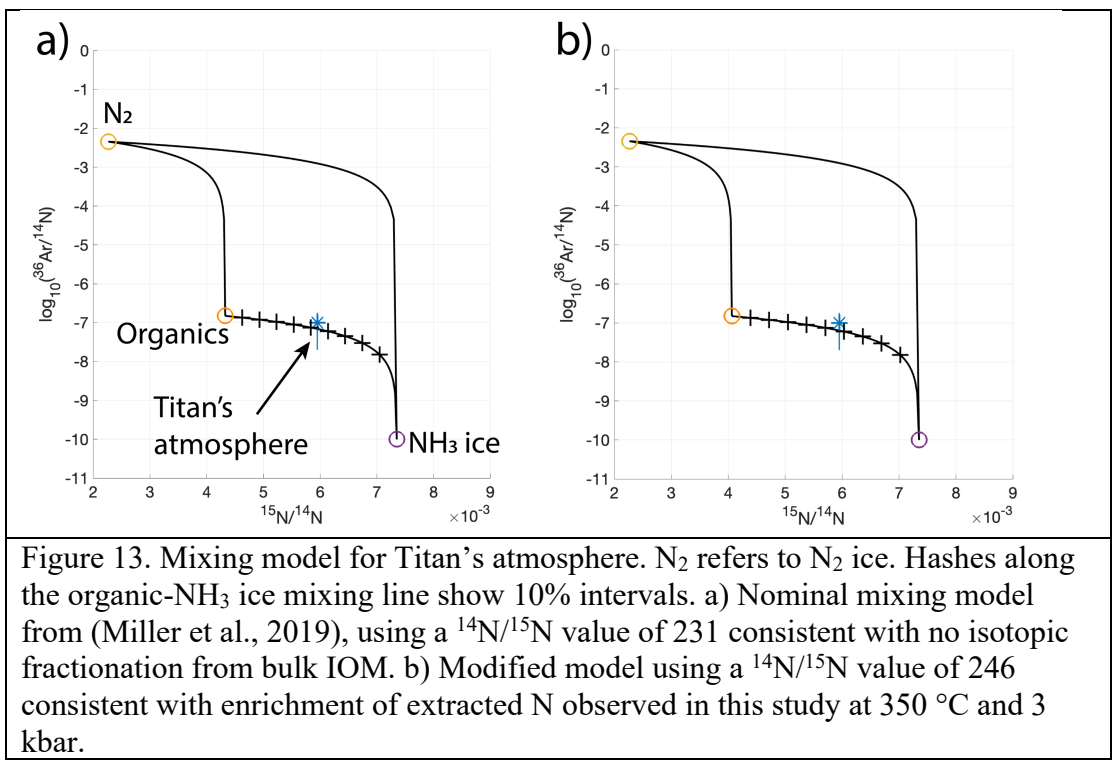

Figure 13. Mixing model for Titan's atmosphere. $N_2$ refers to $N_2$ ice. Hashes along the organic-$NH_3$ ice mixing line show 10% intervals. a) Nominal mixing model from (Miller et al., 2019), using a $^{14}N/^{15}N$ value of 231 consistent with no isotopic fractionation from bulk IOM. b) Modified model using a $^{14}N/^{15}N$ value of 246 consistent with enrichment of extracted N observed in this study at 350 °C and 3 kbar.

## 4.1   Implications for other Ocean Worlds

While our experiments focused on temperatures and pressures relevant to Titan, our results are relevant for other Ocean Worlds. Geophysical constraints suggest that Ceres and Enceladus may have higher abundances of complex organic matter in their interiors than Titan, while Ganymede and especially Europa may have lower abundances (Reynard and Sotin, 2023). Volatilization of such material may be a potential source for $CO_2$ ices observed at the surfaces of Ocean Worlds such as Europa (Trumbo and Brown, 2023; Villanueva et al., 2023) and the Uranian satellites (Cartwright et al., 2015, 2022; Grundy et al., 2003, 2006). To date, such observations have not identified features associated with organic materials at Europa or the Uranian satellites; however, data from Juno's Jovian Infrared Auroral Mapper (JIRAM) instrument suggest the presence of organic and ammoniated compounds at Ganymede's surface (Tosi et al., 2024). Pressures at the water (or high-pressure ice) interface with the rocky interior are generally lower for other Ocean Worlds compared to Titan. For example, the water-rock interface at Europa may be at a pressure of approximately 1.8 kbar (Schubert et al., 2009), while the ocean floor of Enceladus may be only 74 bar (Glein et al., 2018). Relevant interior conditions for these smaller bodies may correspond more closely to our lower pressure experiments. Since $\varepsilon$ remains lower at low pressures over the relevant temperature range, we may expect somewhat more efficient extraction of volatiles like $H_2$ and $CH_4$ from nonpolar regions of the IOM, even at lower temperatures. Volatile production from IOM on other Ocean Worlds may therefore include the production of $CH_4$, even at lower interior temperatures. Future missions such as Europa Clipper and Juice should search for such features in the Jovian system to better understand the origin of $CO_2$ at the surface.

## 5 Conclusions

Heating of complex organic matter similar to chondritic IOM in the presence of water may be an important source of $CH_4$ and $NH_3$ for Titan's atmosphere. Sufficient abundances of both compounds are produced to meet existing constraints on total abundance, outgassing efficiency, and isotopic composition. Volatile production varies as a function of pressure, temperature, and starting composition of the organic matter. Equilibration of carbon-bearing products subsequent to heating and prior to outgassing favors production of $CH_4$. Nitrogen is likely volatilized in the form of $NH_3$, and if IOM is a significant source than subsequent alteration to $N_2$ either in the interior or at Titan's surface is likely required.

## 6 Data Availability

Data are available through Mendeley Data at https://doi.org/10.17632/ztpvzc6kkd.1.

## 7 CRediT Author Statement

**Kelly Miller:** Conceptualization, Investigation, Writing – Original Draft, Visualization, Formal Analysis, **Dionysis Foustoukos:** Methodology, Investigation, Data Curation, Writing – Review & Editing, Validation, **George Cody:** Resources, **Conel Alexander:** Resources, Writing – Review & Editing

## 8 Acknowledgments.

We gratefully acknowledge funding from NASA Solar System Workings grant 80NSSC19K0559, the Emerging Worlds grants 80NSSC20K0344/80NSSC21K0654 (DF, CA, GC), and Exobiology grant 80NSSC21K048 (DF, CA, GC). We thank the Smithsonian National Museum of Natural History for the Murchison meteorite sample used for this study. We also thank two anonymous reviewers for their constructive feedback.

## 9 References

Alexander, C. O'D., Cody, G., De Gregorio, B., Nittler, L., Stroud, R., 2017. The nature, origin and modification of insoluble organic matter in chondrites, the major source of Earth's C and N. Chemie der Erde-Geochemistry 77(227).

Alexander, C.M.O'D., Fogel, M., Yabuta, H., Cody, G., 2007. The origin and evolution of chondrites recorded in the elemental and isotopic compositions of their macromolecular organic matter. Geochimica et Cosmochimica Acta 71(17), 4380-4403.

Alibert, Y., Mousis, O., 2007. Formation of Titan in Saturn's subnebula: constraints from Huygens probe measurements. Astronomy & Astrophysics 465(3), 1051-1060.

Altwegg, K., Balsiger, H., Hänni, N., Rubin, M., Schuhmann, M., Schroeder, I., Sémon, T., Wampfler, S., Berthelier, J.-J., Briois, C., 2020. Evidence of ammonium salts in comet 67P as explanation for the nitrogen depletion in cometary comae. Nature Astronomy 4(5), 533-540.

Atreya, S.K., Adams, E.Y., Niemann, H.B., Demick-Montelara, J.E., Owen, T.C., Fulchignoni, M., Ferri, F., Wilson, E.H., 2006. Titan's methane cycle. Planetary and Space Science 54(12), 1177-1187.

Atreya, S.K., Donahue, T.M., Kuhn, W.R., 1978. Evolution of a nitrogen atmosphere on Titan. Science 201(4356), 611-613.

Atreya, S.K., Lorenz, R.D., Waite, J.H., 2010. Volatile Origin and Cycles: Nitrogen and Methane, in: Brown, R.H., Lebreton, J.-P., Waite, J.H. (Eds.), Titan from Cassini-Huygens. p. 177.

Awolayo, A.N., Tutolo, B.M., 2022. PyGeochemCalc: A Python package for geochemical thermodynamic calculations from ambient to deep Earth conditions. Chemical Geology 606, 120984.

Balsiger, H., Altwegg, K., Bar-Nun, A., Berthelier, J.-J., Bieler, A., Bochsler, P., Briois, C., Calmonte, U., Combi, M., De Keyser, J., 2015. Detection of argon in the coma of comet 67P/Churyumov-Gerasimenko. Science advances 1(8), e1500377.

Bardyn, A., Baklouti, D., Cottin, H., Fray, N., Briois, C., Paquette, J., Stenzel, O., Engrand, C., Fischer, H., Hornung, K., 2017. Carbon-rich dust in comet 67P/Churyumov-Gerasimenko measured by COSIMA/Rosetta. Monthly Notices of the Royal Astronomical Society 469(Suppl_2), S712-S722.

Bethke, C. M. 2022, Geochemical and biogeochemical reaction modeling (Cambridge university press)

Bockelée-Morvan, D., Biver, N., 2017. The composition of cometary ices. Philosophical Transactions of the Royal Society A: Mathematical, Physical and Engineering Sciences 375(2097), 20160252.

Burnham, A. K. 2017, Global Chemical Kinetics of Fossil Fuels: How to Model Maturation and Pyrolysis (Springer)

Burnham, A.K., Sweeney, J.J., 1989. A chemical kinetic model of vitrinite maturation and reflectance. Geochimica et Cosmochimica Acta 53(10), 2649-2657.

Busemann, H., Baur, H., Wieler, R., 2000. Primordial noble gases in "phase Q" in carbonaceous and ordinary chondrites studied by closed-system stepped etching. Meteoritics & Planetary Science 35(5), 949-973.

Canup, R.M., Ward, W.R., 2006. A common mass scaling for satellite systems of gaseous planets. Nature 441(7095), 834-839.

Cartwright, R.J., Emery, J.P., Rivkin, A.S., Trilling, D.E., Pinilla-Alonso, N., 2015. Distribution of $CO_2$ ice on the large moons of Uranus and evidence for compositional stratification of their near-surfaces. Icarus 257, 428-456.

Cartwright, R.J., Nordheim, T.A., DeColibus, R.A., Grundy, W.M., Holler, B.J., Beddingfield, C.B., Sori, M.M., Lucas, M.P., Elder, C.M., Regoli, L.H., Cruikshank, D.P., Emery, J.P., Leonard, E.J., Cochrane, C.J., 2022. A $CO_2$ Cycle on Ariel? Radiolytic Production and Migration to Low-latitude Cold Traps. The Planetary Science Journal 3, 8.

Castillo-Rogez, J.C., Lunine, J.I., 2010. Evolution of Titan's rocky core constrained by Cassini observations. Geophysical Research Letters 37(20), L20205.

Cody, G.D., Alexander, C.M.O.D., Foustoukos, D.I., Busemann, H., Eckley, S., Burton, A.S., Berger, E.L., Nuevo, M., Sandford, S.A., Glavin, D.P., 2024. The nature of insoluble organic matter in Sutter's Mill and Murchison carbonaceous chondrites: Testing the effect of x-ray computed tomography and exploring parent body organic molecular evolution. Meteoritics & Planetary Science 59(1), 3-22.

Cramer, B., 2004. Methane generation from coal during open system pyrolysis investigated by isotope specific, Gaussian distributed reaction kinetics. Organic Geochemistry 35(4), 379-392.

De Sanctis, M., Ammannito, E., Raponi, A., Marchi, S., McCord, T., McSween, H., Capaccioni, F., Capria, M., Carrozzo, F., Ciarniello, M.J.N., 2015. Ammoniated phyllosilicates with a likely outer solar system origin on (1) Ceres. 528(7581), 241.

Derenne, S., Robert, F., 2010. Model of molecular structure of the insoluble organic matter isolated from Murchison meteorite. Meteoritics & Planetary Science 45(9), 1461-1475.

Diamond, L.W., Akinfiev, N.N., 2003. Solubility of CO2 in water from− 1.5 to 100 C and from 0.1 to 100 MPa: evaluation of literature data and thermodynamic modelling. Fluid phase equilibria 208(1-2), 265-290.

Dick, J.M., 2019. CHNOSZ: Thermodynamic calculations and diagrams for geochemistry. Frontiers in Earth Science 7, 180.

Duan, Z., Mao, S., 2006. A thermodynamic model for calculating methane solubility, density and gas phase composition of methane-bearing aqueous fluids from 273 to 523 K and from 1 to 2000 bar. Geochimica et Cosmochimica Acta 70(13), 3369-3386.

Erkaev, N., Scherf, M., Thaller, S., Lammer, H., Mezentsev, A., Ivanov, V., Mandt, K., 2021. Escape and evolution of Titan's N2 atmosphere constrained by 14N/15N isotope ratios. Monthly Notices of the Royal Astronomical Society 500(2), 2020-2035.

Fernandez, D., Goodwin, A., Lemmon, E.W., Levelt Sengers, J., Williams, R., 1997. A formulation for the static permittivity of water and steam at temperatures from 238 K to 873 K at pressures up to 1200 MPa, including derivatives and Debye–Hückel coefficients. Journal of Physical and Chemical Reference Data 26(4), 1125-1166.

Foustoukos, D.I., 2012. Metastable equilibrium in the CHO system: Graphite deposition in crustal fluids. American Mineralogist 97(8-9), 1373-1380.

Foustoukos, D.I., Alexander, C.M.O.D., Cody, G.D., 2021. H and N systematics in thermally altered chondritic insoluble organic matter: An experimental study. Geochimica et Cosmochimica Acta 300, 44.

Foustoukos, D.I., Houghton, J.L., Seyfried, W.E., Jr., Sievert, S.M., Cody, G.D., 2011. Kinetics of $H_2$-$O_2$-$H_2O$ redox equilibria and formation of metastable $H_2O_2$ under low temperature hydrothermal conditions. Geochimica et Cosmochimica Acta 75, 1594-1607.

Foustoukos, D.I., Mysen, B.O., 2015. The structure of water-saturated carbonate melts. American Mineralogist 100, 35-46.

Foustoukos, D.I., Pester, N.J., Ding, K., Seyfried, W.E., 2009. Dissolved carbon species in associated diffuse and focused flow hydrothermal vents at the Main Endeavour Field, Juan de Fuca Ridge: Phase equilibria and kinetic constraints. Geochemistry, Geophysics, Geosystems 10(10).

Glavin, D., Alexander, C., Aponte, J., Dworkin, J., Elsila, J., Yabuta, H., 2018. The Origin and Evolution of Organic Matter in Carbonaceous Chondrites and Links to Their Parent Bodies.

Glein, C., Postberg, F., Vance, S., 2018. The Geochemistry of Enceladus: Composition and Controls. Enceladus and the Icy Moons of Saturn, 39.

Glein, C.R., 2015. Noble gases, nitrogen, and methane from the deep interior to the atmosphere of Titan. Icarus 250, 570-586.

Glein, C.R., Desch, S.J., Shock, E.L., 2009. The absence of endogenic methane on Titan and its implications for the origin of atmospheric nitrogen. Icarus 204(2), 637-644.

Greenberg, J.M., 1986. Predicting that comet Halley is dark. Nature 321, 385.

Grundy, W.M., Young, L.A., Spencer, J.R., Johnson, R.E., Young, E.F., Buie, M.W., 2006. Distributions of $H_2O$ and $CO_2$ ices on Ariel, Umbriel, Titania, and Oberon from IRTF/SpeX observations. Icarus 184, 543-555.

Grundy, W.M., Young, L.A., Young, E.F., 2003. Discovery of $CO_2$ ice and leading-trailing spectral asymmetry on the uranian satellite ariel. Icarus 162, 222-229.

Hashizume, K., Ishida, A., Chiba, A., Okazaki, R., Yogata, K., Yada, T., Kitajima, F., Yurimoto, H., Nakamura, T., Noguchi, T., Yabuta, H., Naraoka, H., Takano, Y., Sakamoto, K., Tachibana, S., Nishimura, M., Nakato, A., Miyazaki, A., Abe, M., Okada, T., Usui, T., Yoshikawa, M., Saiki, T., Terui, F., Tanaka, S., Nakazawa, S., Watanabe, S.-i., Tsuda, Y., Broadley, M.W., Busemann, H., Team, the Hayabusa2 Initial Analysis Volatile Team, 2024. The Earth atmosphere-like bulk nitrogen isotope composition obtained by stepwise combustion analyses of Ryugu return samples. Meteoritics & Planetary Science.

Heger, K., Uematsu, M., Franck, E., 1980. The static dielectric constant of water at high pressures and temperatures to 500 MPa and 550 C. Berichte der Bunsengesellschaft für physikalische Chemie 84(8), 758-762.

Holmes, R., McClelland, J., Sigman, D., Fry, B., Peterson, B., 1998. Measuring 15N–NH4+ in marine, estuarine and fresh waters: an adaptation of the ammonia diffusion method for samples with low ammonium concentrations. Marine Chemistry 60(3-4), 235-243.

Hopp, T., Dauphas, N., Abe, Y., Aléon, J., O'D. Alexander, C.M.O.D., Amari, S., Amelin, Y., Bajo, K.-i., Bizzarro, M., Bouvier, A., 2022. Ryugu's nucleosynthetic heritage from the outskirts of the Solar System. Science advances 8(46), eadd8141.

Hörst, S.M., Vuitton, V., Yelle, R.V., 2008. Origin of oxygen species in Titan's atmosphere. Journal of Geophysical Research: Planets 113(E10).

Idini, B., Nimmo, F., 2024. Resonant Stratification in Titan's Global Ocean. The Planetary Science Journal 5(1), 15.

Johnson, J.W., Norton, D., 1991. Critical phenomena in hydrothermal systems; state, thermodynamic, electrostatic, and transport properties of H 2 O in the critical region. American Journal of Science 291(6), 541-648.

Kalousová, K., Sotin, C., 2020. The insulating effect of methane clathrate crust on Titan's thermal evolution. Geophysical Research Letters 47(13), e2020GL087481.

Kissel, J., Brownlee, D., Büchler, K., Clark, B., Fechtig, H., Grün, E., Hornung, K., Igenbergs, E., Jessberger, E., Krueger, F., 1986. Composition of comet Halley dust particles from Giotto observations. Nature 321(6067), 336-337.

Kissel, J., Krueger, F., 1987. The organic component in dust from comet Halley as measured by the PUMA mass spectrometer on board Vega 1. Nature 326, 755-760.

Kung, C.-C., Clayton, R.N., 1978. Nitrogen abundances and isotopic compositions in stony meteorites. Earth and Planetary Science Letters 38(2), 421-435.

Leitner, M.A., Lunine, J.I., 2019. Modeling early Titan's ocean composition. Icarus 333, 61-70.

Lewis, J., Prinn, R., 1980. Kinetic inhibition of CO and N2 reduction in the solar nebula. The Astrophysical Journal 238, 357-364.

Mandt, K.E., Mousis, O., Lunine, J., Gautier, D., 2014. Protosolar ammonia as the unique source of Titan's nitrogen. The Astrophysical Journal Letters 788(2), L24.

Mandt, K.E., Waite, J.H., Lewis, W., Magee, B., Bell, J., Lunine, J., Mousis, O., Cordier, D., 2009. Isotopic evolution of the major constituents of Titan's atmosphere based on Cassini data. Planetary and Space Science 57(14), 1917-1930.

Mandt, K.E., Waite, J.H., Teolis, B., Magee, B.A., Bell, J., Westlake, J.H., Nixon, C.A., Mousis, O., Lunine, J.I., 2012. The 12C/13C ratio on Titan from Cassini INMS measurements and implications for the evolution of methane. The Astrophysical Journal 749(2), 160.

Marounina, N., Tobie, G., Carpy, S., Monteux, J., Charnay, B., Grasset, O., 2015. Evolution of Titan's atmosphere during the Late Heavy Bombardment. Icarus 257, 324-335.

Mastalerz, M., Wilks, K., Bustin, R., Ross, J., 1993. The effect of temperature, pressure and strain on carbonization in high-volatile bituminous and anthracitic coals. Organic Geochemistry 20(2), 315-325.

McDonnell, J., Alexander, W., Burton, W., Bussoletti, E., Clark, D., Grard, R., Grün, E., Hanner, M., Hughes, D., Igenbergs, E., 1986. Dust density and mass distribution near comet Halley from Giotto observations. Nature 321(6067), 338-341.

McKay, C.P., Scattergood, T.W., Pollack, J.B., Borucki, W.J., Van Ghyseghem, H.T., 1988. High-temperature shock formation of N2 and organics on primordial Titan. Nature 332(7), 520-522.

Melwani Daswani, M., Vance, S.D., Mayne, M.J., Glein, C.R., 2021. A metamorphic origin for Europa's ocean. Geophysical Research Letters 48(18), e2021GL094143.

Miller, K.E., Glein, C.R., Waite Jr, J.H., 2019. Contributions from accreted organics to Titan's atmosphere: New insights from cometary and chondritic data. Astrophysical Journal 871, 59.

Mitri, G., Meriggiola, R., Hayes, A., Lefevre, A., Tobie, G., Genova, A., Lunine, J.I., Zebker, H., 2014. Shape, topography, gravity anomalies and tidal deformation of Titan. Icarus 236, 169-177.

Mousis, O., Lunine, J.I., Thomas, C., Pasek, M., Marbœuf, U., Alibert, Y., Ballenegger, V., Cordier, D., Ellinger, Y., Pauzat, F., 2009. Clathration of volatiles in the solar nebula and implications for the origin of Titan's atmosphere. The Astrophysical Journal 691(2), 1780.

Niemann, H., Atreya, S., Bauer, S., Carignan, G., Demick, J., Frost, R., Gautier, D., Haberman, J., Harpold, D., Hunten, D., 2005. The abundances of constituents of Titan's atmosphere from the GCMS instrument on the Huygens probe. Nature 438(7069), 779-784.

Niemann, H.B., Atreya, S.K., Demick, J.E., Gautier, D., Haberman, J.A., Harpold, D.N., Kasprzak, W.T., Lunine, J.I., Owen, T.C., Raulin, F., 2010. Composition of Titan's lower atmosphere and simple surface volatiles as measured by the Cassini-Huygens probe gas chromatograph mass spectrometer experiment. Journal of Geophysical Research: Planets 115(E12), E12006.

Nixon, C., Temelso, B., Vinatier, S., Teanby, N., Bézard, B., Achterberg, R., Mandt, K., Sherrill, C., Irwin, P., Jennings, D.J.T.A.J., 2012. Isotopic ratios in Titan's methane: measurements and modeling. 749(2), 159.

Oba, Y., Naraoka, H., 2009. Elemental and isotope behavior of macromolecular organic matter from CM chondrites during hydrous pyrolysis. Meteoritics & Planetary Science 44(7), 943-953.

Ott, U., Mack, R., Sherwood, C., 1981. Noble-gas-rich separates from the Allende meteorite. Geochimica et Cosmochimica Acta 45(10), 1751-1788.

Owen, T., 1982. The composition and origin of Titan's atmosphere. Planetary and Space Science 30(8), 833-838.

Pérez-Rodríguez, I., Sievert, S.M., Fogel, M.L., Foustoukos, D.I., 2017. Biogeochemical N signatures from rate-yield trade-offs during in vitro chemosynthetic $NO_3^-$ reduction by deep-sea vent ε-Proteobacteria and Aquificae growing at different temperatures. Geochimica et Cosmochimica Acta 211, 214-227.

Pizzarello, S., Feng, X., Epstein, S., Cronin, J., 1994. Isotopic analyses of nitrogenous compounds from the Murchison meteorite: ammonia, amines, amino acids, and polar hydrocarbons. Geochimica et Cosmochimica Acta 58(24), 5579-5587.

Pizzarello, S., Williams, L., 2012. Ammonia in the early solar system: An account from carbonaceous meteorites. The Astrophysical Journal 749(2), 161.

Poch, O., Istiqomah, I., Quirico, E., Beck, P., Schmitt, B., Theulé, P., Faure, A., Hily-Blant, P., Bonal, L., Raponi, A., 2020. Ammonium salts are a reservoir of nitrogen on a cometary nucleus and possibly on some asteroids. Science 367(6483), eaaw7462.

Prinn, R.G., Fegley, B., Jr., 1981. Kinetic inhibition of CO and N2 reduction in circumplanetary nebulae - Implications for satellite composition. The Astrophysical Journal 249, 308.

Remusat, L., Derenne, S., Robert, F., Knicker, H., 2005. New pyrolytic and spectroscopic data on Orgueil and Murchison insoluble organic matter: A different origin than soluble? Geochimica et Cosmochimica Acta 69(15), 3919-3932.

Reynard, B., Sotin, C., 2023. Carbon-rich icy moons and dwarf planets. Earth and Planetary Science Letters 612, 118172.

Rousselot, P., Pirali, O., Jehin, E., Vervloet, M., Hutsemékers, D., Manfroid, J., Cordier, D., Martin-Drumel, M.-A., Gruet, S., Arpigny, C., 2013. Toward a unique nitrogen isotopic ratio in cometary ices. The Astrophysical Journal Letters 780(2), L17.

Safi, E., Thompson, S.P., Evans, A., Day, S.J., Murray, C., Parker, J., Baker, A., Oliveira, J., Van Loon, J.T., 2017. Properties of $CO_2$ clathrate hydrates formed in the presence of $MgSO_4$ solutions with implications for icy moons. Astronomy & Astrophysics 600, A88.

Schubert, G., Sohl, F., Hussmann, H., 2009. Interior of europa. Europa, 353-367.

Sekine, Y., Genda, H., Sugita, S., Kadono, T., Matsui, T., 2011. Replacement and late formation of atmospheric $N_2$ on undifferentiated Titan by impacts. Nature Geoscience 4(6), 359-362.

Sessions, A.L., 2001. Hydrogen isotope ratios of individual organic compounds. Indiana University, Indiana.

Shinnaka, Y., Kawakita, H., Jehin, E., Decock, A., Hutsemékers, D., Manfroid, J., Arai, A., 2016. Nitrogen isotopic ratios of $NH_2$ in comets: implication for 15N-fractionation in cometary ammonia. Monthly Notices of the Royal Astronomical Society 462(Suppl_1), S195-S209.

Shinnaka, Y., Kawakita, H., Kobayashi, H., Nagashima, M., Boice, D.C., 2014. $14NH_2/15NH_2$ ratio in comet C/2012 S1 (ISON) observed during its outburst in 2013 November. The Astrophysical Journal Letters 782(2), L16.

Sigman, D., Altabet, M., Michener, R., McCorkle, D., Fry, B., Holmes, R., 1997. Natural abundance-level measurement of the nitrogen isotopic composition of oceanic nitrate: an adaptation of the ammonia diffusion method. Marine chemistry 57(3-4), 227-242.

Sloan, E., Koh, C., 2007. Molecular structures and similarities to ice. Clathrate hydrates of natural gases 3.

Sotin, C., Kalousová, K., Tobie, G., 2021. Titan's Interior Structure and Dynamics After the Cassini-Huygens Mission. Annual Review of Earth and Planetary Sciences 49, 579-607.

Strobel, D.F., Shemansky, D., 1982. EUV emission from Titan's upper atmosphere: Voyager 1 encounter. Journal of Geophysical Research: Space Physics 87(A3), 1361-1368.

Sverjensky, D.A., Harrison, B., Azzolini, D., 2014. Water in the deep Earth: the dielectric constant and the solubilities of quartz and corundum to 60 kb and 1200 C. Geochimica et Cosmochimica Acta 129, 125-145.

Tobie, G., Gautier, D., Hersant, F., 2012. Titan's Bulk Composition Constrained by Cassini-Huygens: Implication for Internal Outgassing. The Astrophysical Journal 752(2), 125.

Tobie, G., Lunine, J.I., Sotin, C., 2006. Episodic outgassing as the origin of atmospheric methane on Titan. Nature 440(7080), 61.

Tosi, F., Mura, A., Cofano, A., Zambon, F., Glein, C.R., Ciarniello, M., Lunine, J.I., Piccioni, G., Plainaki, C., Sordini, R., 2024. Salts and organics on Ganymede's surface observed by the JIRAM spectrometer onboard Juno. Nature Astronomy 8(1), 82-93.

Trumbo, S.K., Brown, M.E., 2023. The distribution of $CO_2$ on Europa indicates an internal source of carbon. Science 381(6664), 1308-1311.

van Krevelen, D.W., 1982. Development of coal research—a review. Fuel 61(9), 786-790.

Villanueva, G., Hammel, H., Milam, S., Faggi, S., Kofman, V., Roth, L., Hand, K., Paganini, L., Stansberry, J., Spencer, J., 2023. Endogenous $CO_2$ ice mixture on the surface of Europa and no detection of plume activity. Science 381(6664), 1305-1308.

Vollmer, C., Kepaptsoglou, D., Leitner, J., Mosberg, A.B., El Hajraoui, K., King, A.J., Bays, C.L., Schofield, P.F., Araki, T., Ramasse, Q.M., 2024. High-spatial resolution functional chemistry of nitrogen compounds in the observed UK meteorite fall Winchcombe. Nature Communications 15(1), 778.

Waite, J.H., Niemann, H., Yelle, R.V., Kasprzak, W.T., Cravens, T.E., Luhmann, J.G., McNutt, R.L., Ip, W.-H., Gell, D., De La Haye, V., 2005. Ion neutral mass spectrometer results from the first flyby of Titan. Science 308(5724), 982-986.

Yabuta, H., Cody, G.D., Engrand, C., Kebukawa, Y., De Gregorio, B., Bonal, L., Remusat, L., Stroud, R., Quirico, E., Nittler, L., 2023. Macromolecular organic matter in samples of the asteroid (162173) Ryugu. Science 379(6634), eabn9057.

Yabuta, H., Williams, L.B., Cody, G.D., Alexander, C.M.O.D., Pizzarello, S., 2007. The insoluble carbonaceous material of CM chondrites: A possible source of discrete organic compounds under hydrothermal conditions. Meteoritics & Planetary Science 42(1), 37-48.

Yu, X., Thompson, M., Duncan, T., Kim, K., Telus, M., Toyanath, J., Lederman, D., Zhang, X., 2022. Carbonaceous Chondrite Outgassing Experiments: Implications for Methane Replenishment on Titan, Lunar and Planetary Science Conference. pp. 2899.

Yung, Y.L., Allen, M., Pinto, J.P., 1984. Photochemistry of the atmosphere of Titan-Comparison between model and observations. The Astrophysical Journal Supplement Series 55, 465-506.